\documentclass[prepreint,twocolumn]{aastex62}
\hypersetup{linkcolor={blue},citecolor={blue}}
\usepackage{bm}
\usepackage{url}
\usepackage{color}
\usepackage{soul}
\usepackage{ulem} 
\usepackage{amsmath}
\usepackage{float}
\usepackage{amssymb}
\usepackage{natbib}
\usepackage{xspace}
\usepackage{graphicx}
\usepackage{rotating}
\usepackage{multirow}
\usepackage{graphicx}
\usepackage{subfigure}
\def\psqcm{cm$^{-2}$}
\def\H0 {{\it H}$_0$}

\def\q0 {{\it q}$_0$}

\def\ergps{erg~s$^{-1}$}

\def\nH{$N_{\rm H}$\thinspace} 
\def\psqcm{cm$^{-2}$} 

\def\phpspsqcm{photons \thinspace cm$^{-2}$ \thinspace s$^{-1}$}
\newcommand{\nh} {$N_{\text{H}}$}

\begin{document}
 
\title{The kpc Scale Fe K$\alpha$ Emission in the  Compton Thin Seyfert 2 Galaxy NGC 4388 resolved by Chandra}

\correspondingauthor{J. Wang}
\email{jfwang@xmu.edu.cn}
\author{Huili Yi}
\affiliation{Department of Astronomy, Xiamen University, Xiamen, Fujian 361005, China}
\author{Junfeng Wang}
\affiliation{Department of Astronomy, Xiamen University, Xiamen, Fujian 361005, China}
\author{Xinwen Shu}
\affiliation{Department of Physics, Anhui Normal University, Wuhu, Anhui, 241000, China}
\author{Giuseppina Fabbiano}
\affil{Center for Astrophysics, Harvard \& Smithsonian, 60 Garden St. Cambridge, MA 02138, USA}
\author{Cirino Pappalardo}
\affiliation{Instituto de Astrof\'{i}sica e Ci\^{e}ncias do Espa\c{c}o, Universidade de Lisboa - OAL, Tapada da Ajuda, PT1349-018 Lisboa, Portugal}
\affiliation{Departamento de F\'{i}sica, Faculdade de Ci\^{e}ncias da Universidade de Lisboa, Edif\'{i}cio C8, Campo Grande, PT1749-016 Lisboa, Portugal}
\author{Chen Wang}
\affiliation{Department of Astronomy, Xiamen University, Xiamen, Fujian 361005, China}
\author{Hanbo Yu}
\affiliation{Department of Astronomy, Xiamen University, Xiamen, Fujian 361005, China}
\begin{abstract}

We present {\em Chandra} imaging and spectral observations of Seyfert 2 galaxy NGC 4388. Three extended X-ray structures around the nucleus on kpc scale are well imaged, allowing an in-depth spatially resolved study. Both the extended hard continuum and the Fe K$\alpha$ line show similar morphology, consistent with a scenario where the ionizing emission from nucleus is reprocessed by circumnuclear cold gas, resulting in a weak reflection continuum and an associated neutral Fe K$\alpha$ line. This has been seen in other Compton thick active galactic nuclei (AGN), but NGC 4388 is one of the rare cases with a lower column density ($N_{\rm H} < 1.25\times 10^{24}$ cm$^{-2}$) along the line of sight. Significant differences in equivalent width of the Fe K$\alpha$ emission line are found for the nuclear and extended regions, which could be ascribed to different column densities or scattering angles with respect to the line of sight, rather than variations in iron abundances. The north-east and west extended structures are aligned with the galactic disk and dust lane in the HST $V-H$ map, and located at the peak of molecular gas distribution. The morphology implies that the kpc-scale radio outflow may have compressed the interstellar gas and produced clumps working as the reflector to enhance line emission. Using [OIV] emission as a proxy of the AGN intrinsic luminosity, we find that both of the extended Fe K$\alpha$ emission and reflection continuum are linearly correlated with the [OIV] luminosity, indicating a connection between the AGN and the extended emission.

\end{abstract}

\keywords{Galaxies: active -- Galaxies: Seyfert--Individual: NGC 4388}

\section{Introduction}\label{sec:intro}
The majority of nearby active galactic nuclei (AGN) are obscured by a large amount of circumnuclear cold gas and dust \citep {2004Coma}. The primary emission from nucleus is reprocessed by these circumnuclear material, with a resulting Compton reflection continuum and an associated fluorescent Fe K$\alpha$ line at 6.4 keV \citep {Shu2010,1996Matt,1991George,Winter2009,Jiang2006,Levenson2002,Levenson2006,Yaqoob2004,Perola2002,Weaver2001,shu2011}. Indeed, most of our knowledge so far about the circumnuclear reprocessing matter, at least when their X-ray properties are concerned, is based on the brightest Compton thick sources with absorption column densities larger than 1.25$\times10^{24}$ cm$^{-2}$ (the inverse of the Thomson scattering cross-section) due to their heavy obscuration. The nearby Compton thick AGN NGC 1068 \citep{2015Bauer}, NGC 4945 \citep{Marinucci2017}, Circinus \citep{2013Marinucci}, ESO 428-G014 \citep{Fabbiano2017Discovery} and NGC 5643 \citep{Fabbiano2018,Alonso-Herrero2018} are good cases for which clumpy fluorescing clouds around nucleus have been spatially resolved. Compared with the above Compton thick sources, it is more challenging for us to resolve the circumnuclear reflector in a Compton thin source because extended emission is usually overwhelmed by the bright nucleus. In this work, we discover that NGC 4388 which is a Compton thin source (  $N_{\text{H}}$ $\simeq 3.6^{+0.3}_{-0.2} \times 10^{23}$ \psqcm, this work) provides us a new perspective to investigate the properties of extended reprocessing matter.  

NGC 4388 ($z = 0.00842$) is one of the brightest Seyfert 2 galaxies at hard X-ray energies \citep{Baumgartner_2013} with log L$_{20-50}^{int}$ =43.23  $erg~s^{-1}$ \citep{2017Ricci} . Several broadband observations of this source have been performed with X-ray missions such as ASCA \citep{iwasawa1997asca}, INTEGRAL and Swift \citep{fedorova2011studying}, XMM-Newton \citep{2004Beckmann}, BeppoSAX \citep{elvis2004unveiling}, Suzaku \citep{shirai2008detailed}, NuSTAR \citep{2017ApJ...843...89K} supporting the hypothesis of a powerful hidden nucleus. Based on more than 30 years of X-ray observations reported in the literature, the column density in this source shows \nH variability spanning from 2 to 7 $ \times $ $10^{23}$ cm$^{-2}$  which are tabulated in Table 2. Furthermore, a strong neutral Fe K$\alpha$ emission line is detected at 6.4 keV in above observations for the nucleus. This result has also been confirmed by Chandra \citep{2003MN} at an energy of $6.36^{+0.02}_{-0.02}$ keV, with an equivalent width (EW) of $440\pm 90$ eV in agreement with ASCA observation of EW$\sim 500$ eV \citep{iwasawa1997asca} and NuSTAR of EW$ = 368^{+56}_{-53}$ eV, $394^{+131}_{-95}$ eV \citep{2017ApJ...843...89K}. However, these broadband observations data did not allow us to spectrally and spatially characterize the extra-nuclear extended regions with much detail. 

The high angular resolution of Chandra allows such investigation with arcsecond spatial resolution.  With a 19.97 ks Chandra observation (ObsId 1619), \citet{2003MN} mainly investigated the soft extended X-ray emission. For the extended Fe K$\alpha$ emission, only a rough morphology and extent were reported limited by the short exposure of the observation. 
With our new 28 ks observation (ObsId 12291; PI: Wang), the deeper combined Chandra data will allow us to further map the geometrical structure of Fe K$\alpha$ emitting material and spectrally characterize these regions in detail. Also, the fluorescent neutral Fe K$\alpha$ line is a reliable indicator of cold gas. In this paper, we attempt to verify whether the reflecting medium such as higher column density molecular clouds exist in the circumnuclear region of NGC 4388 to reprocess the nuclear radiation and produce strong Fe K$\alpha$ line.
 
The paper is structured as follows: in Section 2, we discuss the merged Chandra observations and data reduction. In Section 3, we present the imaging and spectral analyses, respectively. Finally, we discuss and summarize our results in Sections 4 and 5. The adopted distance of this galaxy is 16.7 Mpc \citep{1997Yasuda}. The corresponding angular scale is $\sim 81$ pc arcsec$^{-1}$. We adopt the cosmological parameters $H_0=70$ km s$^{-1}$ Mpc$^{-1}$, $\Omega_\Lambda=0.73$ and $\Omega_m=0.27$.

\begin{figure*}
	\includegraphics[width=0.71\columnwidth]{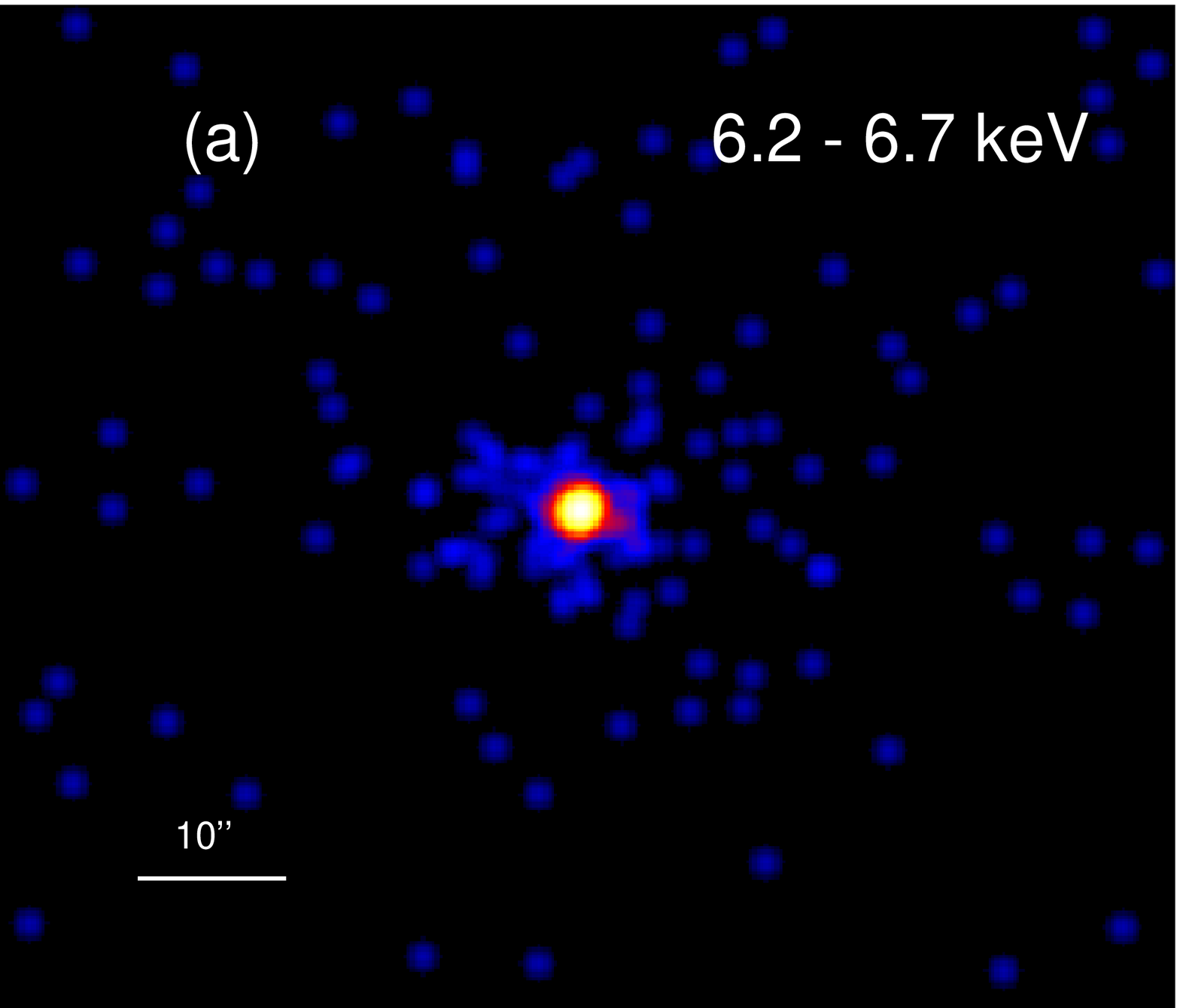}
	\includegraphics[width=0.7\columnwidth]{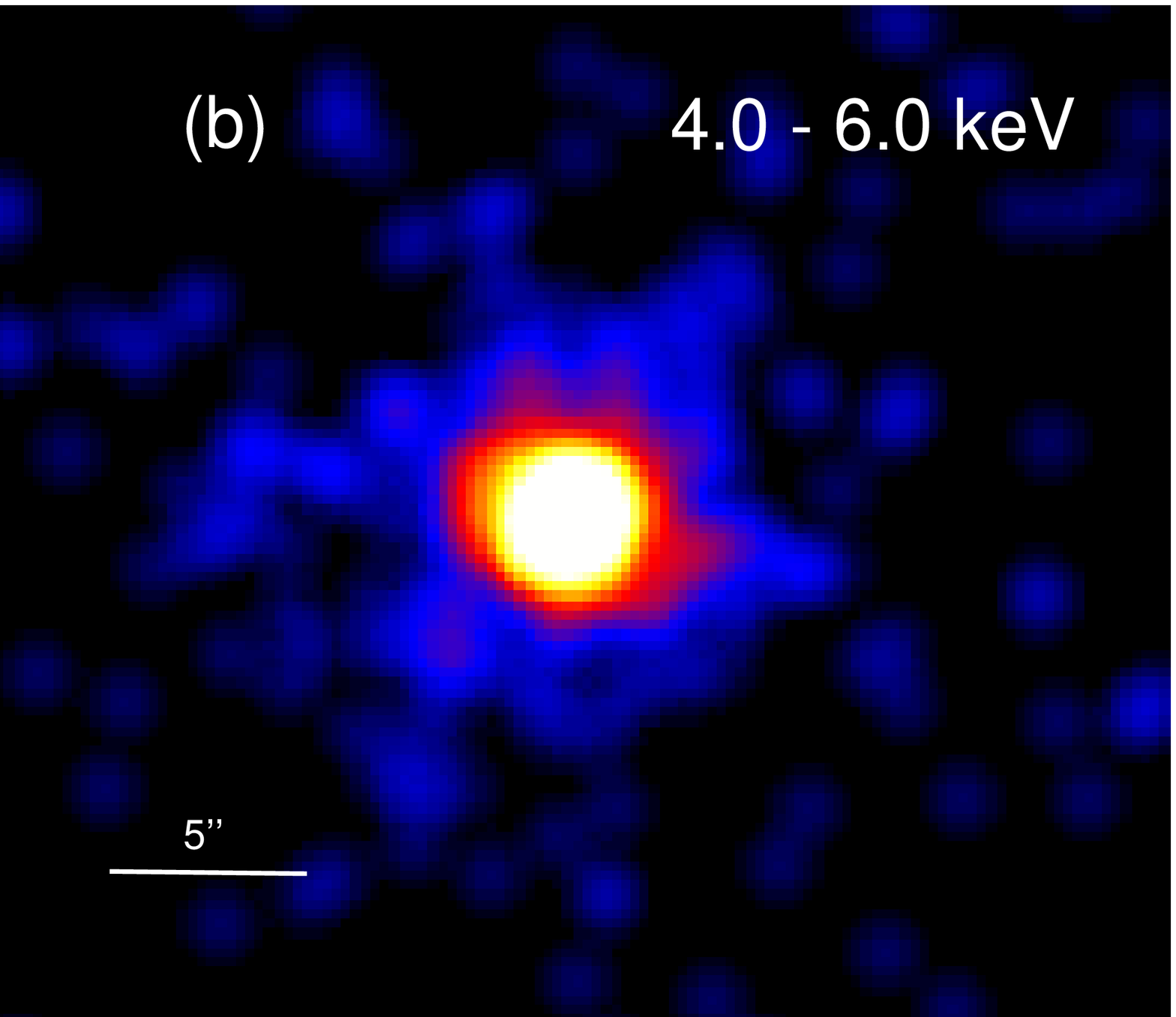}
	\includegraphics[width=0.7\columnwidth]{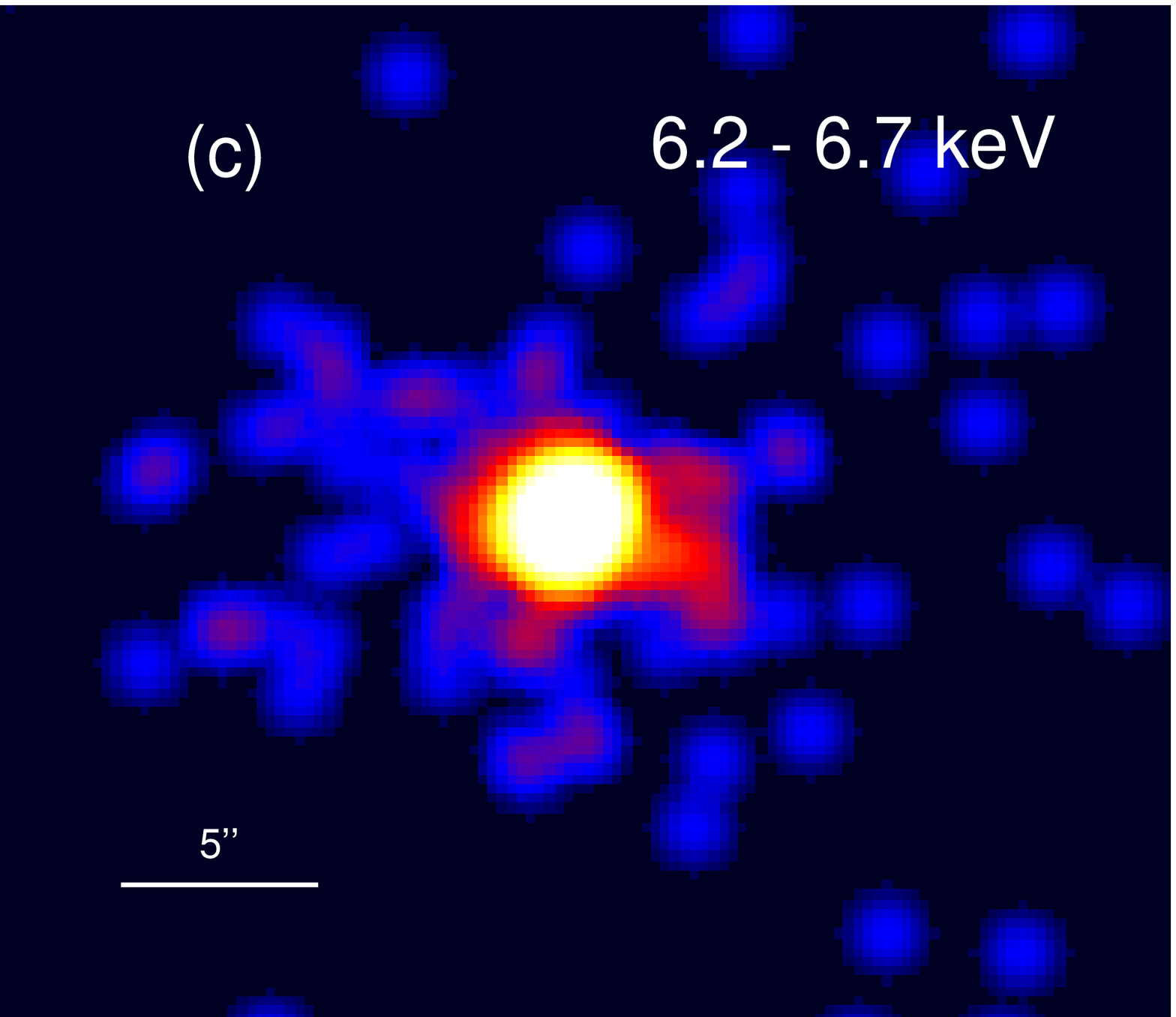}
	\caption{(a) Merged Chandra/ACIS-S image of the central 80$\times$80 arcsec region of NGC 4388, in the 6.2--6.7 keV energy band. (b) Merged ACIS-S image of the central 30$\times$30 arcsec region of NGC 4388, in the 4.0--6.0 keV energy band. (c) ACIS-S image of the central 30$\times$30 arcsec region of NGC 4388 in the 6.2--6.7 keV energy band, shown in a binning of 0.5 native ACIS pixel (the size of a pixel is therefore 0.246 arcsec). The image is smoothed with a Gaussian kernel of 1.25 arcsec.}
	
\end{figure*}

  \section{observations and data reduction}

  NGC 4388 has been observed with the Advanced CCD Imaging Spectrometer \citep[ACIS;][]{Garmire03} onboard Chandra in 2001 and 2011, and with the High Energy Transmission Grating \citep[HETG;][]{Cani05} in 2008. We summarize the four observations of interest in Table 1.   
  
  \begin{table}
  \begin{center}
  \caption{Log of {\it Chandra}/ACIS-S Observations of NGC 4388}
  \begin{tabular}{ccccc}
  \hline
  {\bf Obs. ID} & {\bf Date} & {\bf Exp. Time (ks)} & {\bf Grating }& {\bf Flux$^{\dagger}$} \\	
  \hline
  1619 & 2001-06-08 & 19.97& NONE & 0.87 \\
  9276 & 2008-04-16 & 170.59 & HETG & 1.76 \\
  9277 & 2008-04-24 & 98.31  & HETG & 2.45 \\
  12291 & 2011-12-07 & 27.6  & NONE & 0.60 \\
  \hline
\end{tabular}
\end{center}
$^{\dagger}$ The observed 2.0--10.0 keV flux in units of $10^{-11}$ erg cm$^{-2}$ s$^{-1}$.
\end{table}

As shown in Table 1, the 2--10 keV flux varied by a factor of 4.1 over the past ten years. Flux variations is normal for NGC 4388 in light of previous observational results \citep{1999Forster,2004Beckmann,fedorova2011studying,2007Fukazawa}. \citet{fedorova2011studying} derives a maximum variability in flux by a factor of $\approx$ 2 in the 20-60 keV band by INTEGRAL data on time-scales of 1-2 months while \citet{2004Beckmann} shows the flux variability by a factor of 4 over the year 1993--2002. Previous ASCA observations in 1995 \citep{1999Forster} gives a lower 2-10 keV flux value 0.64$ \times 10^{-11}$ erg~cm$^{-2}$~s$^{-1}$, in line with the ObsIds 1619 and 12291 results. The flux for ObsIds 9276 and 9277 in Table 1 are consistent with the high value 2.31$ \times 10^{-11}$ erg~cm$^{-2}$~s$^{-1}$ provided by \citet{Shu2010}. We first attempted to use the zero-th order image of the deep grating observations (ObsIds 9276 and 9277) to check the Fe K$\alpha$ extended emission, but unfortunately their extended morphologies are dominated by the high luminosity nuclear PSF wings. Thus, we excluded the two HETG observations in the following analysis. 

The remaining two ACIS-S observations (ObsIDs 1619 and 12291) were reduced with the Chandra Interactive Analysis of Observations (CIAO; version 4.9) and the Chandra Calibration Data Base (CALDB; version 4.7.8), adopting standard procedures. We generated event files for the two observations separately with the CIAO tool \textsc{CHANDRA\_REPRO} and merged them into a single image using the tool \textsc{MERGE\_OBS}. The nucleus of this galaxy was positioned on the aimpoint of ACIS-S3 detector.  The background during observations was relatively stable. Brief periods of high background flaring ($\geq3 \sigma$) were filtered out, resulting in a cumulative exposure time of 47.5 ks. The data at the position of nucleus are slightly affected by photon "pile-up", i.e., the phenomena of two or more photons overlapping in a single detector frame and being read out as a single event. The estimated pile-up fraction of NGC 4388 is about 10 percent. We correct these effects following the procedure described in \citet{Davis2001}.

\begin{figure*}
	\centering
	\includegraphics [width=\columnwidth]{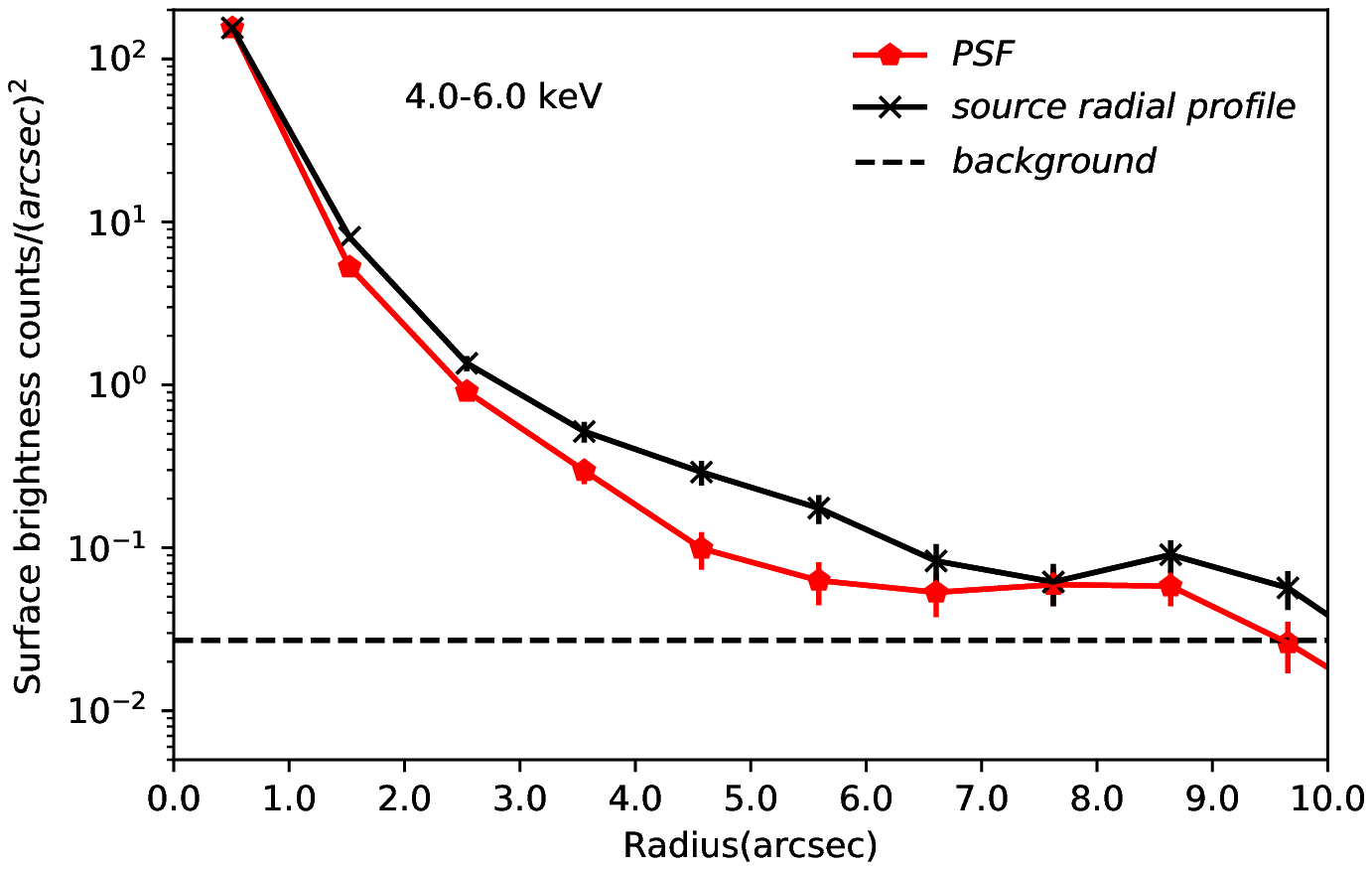}
	\includegraphics [width=\columnwidth]{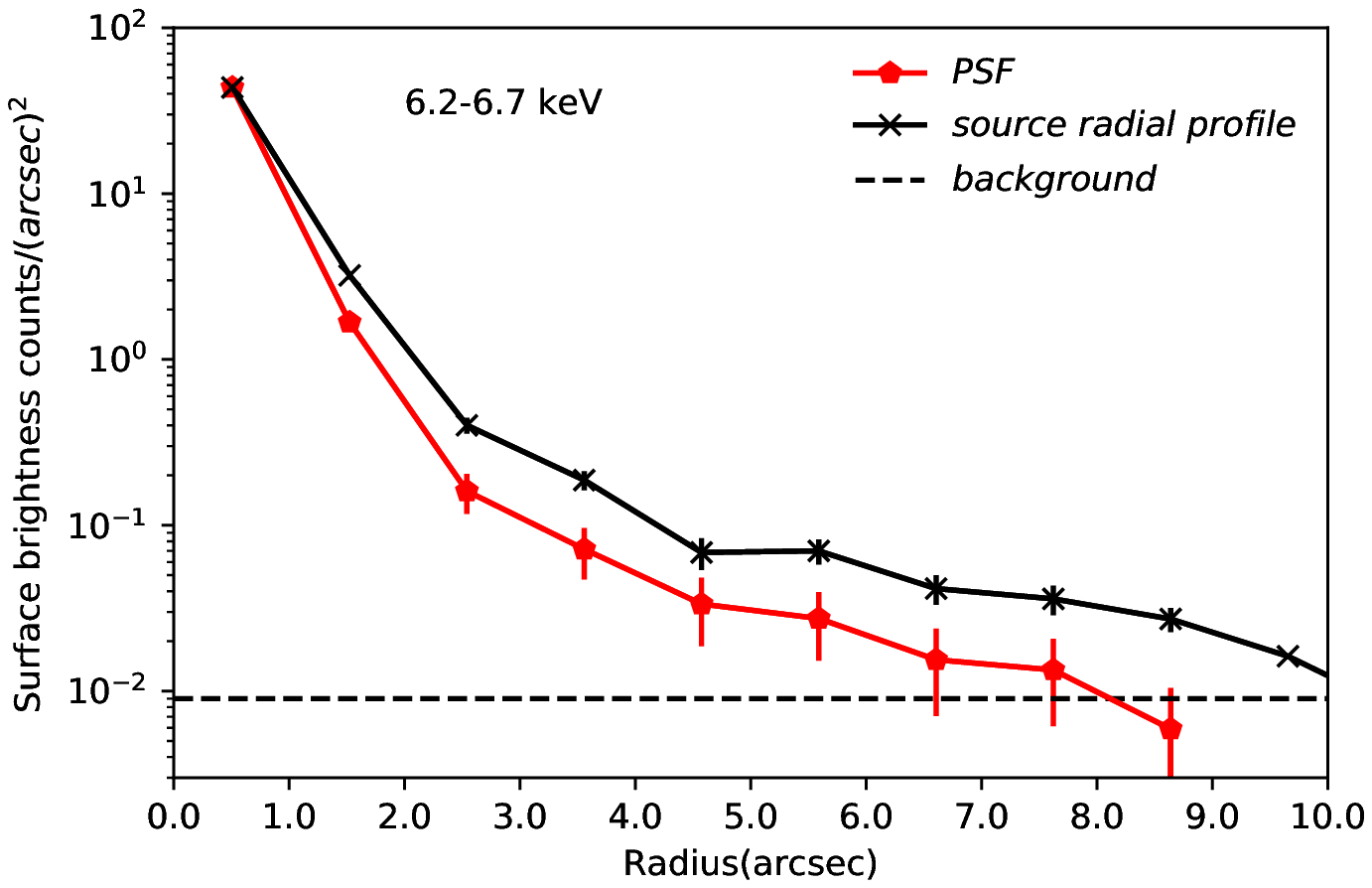}
	\caption{(a) Radial profiles of the source and ACIS-S PSF in the 4.0--6.0 keV energy band. (b) Radial profiles of the source and ACIS-S PSF in 6.2--6.7 keV energy band. We chose 2 pixel radius concentric annuli for the source and PSF surface brightness, and a 30 pixel radius annulus for the background. The PSF radial profile was normalized to the peak of the source.}
\end{figure*}

Spectra from the nuclear source were extracted from a circular region with a radius of 1.5 arcsec (Figure 4). We used a 5 arcsec radius circle for background extraction, located in a nearby source free region, well away from the diffuse emission. The spectra taken from extended regions between 1.5 and 10 arcsecond in radius are shown in Figure 6. Spectra were binned no less than 15 counts in each background-subtracted spectral channel. This allows the applicability of the $\chi^2$ statistics. We performed the spectral modeling using \textsc{XSPEC 12.9.0} \citep{xs}. Errors correspond to 90\% confidence level for parameters ($\Delta\chi^2=2.7$), if not stated otherwise.	  
 
 \begin{figure}[ht]
  \centering
  \includegraphics [width=\columnwidth]{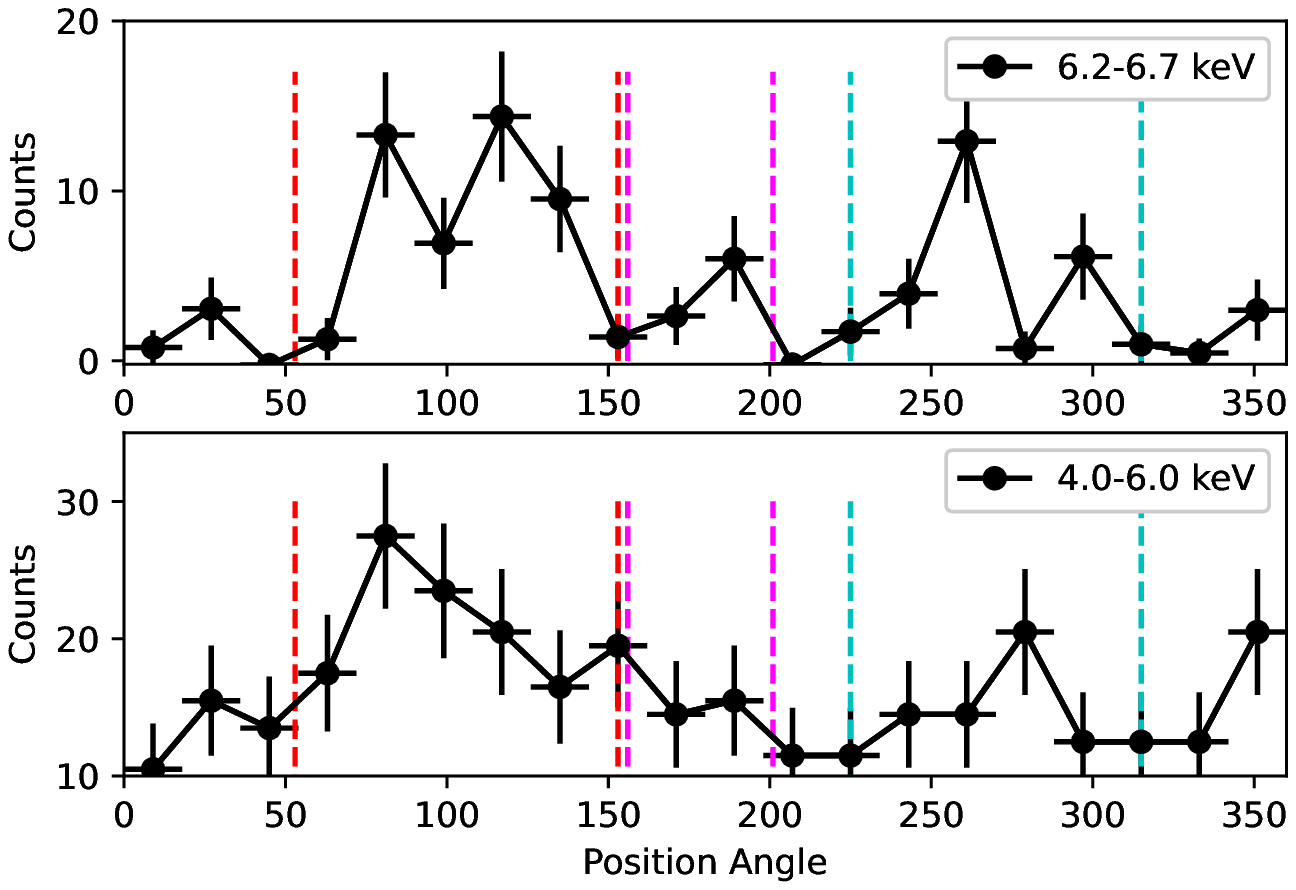}
  \caption{The position angles of sectors that show counts excess in 6.2--6.7 keV and 4.0--6.0 keV (measured counter-clockwise from north), with angular bins of $18^{\circ}$. The east (between red dashed lines) and south (between magenta dashed lines) cones correspond to PA from $53^{\circ}$ to $201^{\circ}$, and the west cone (between cyan dashed lines) corresponds to PA from $225^{\circ}$ to $315^{\circ}$.}
  \end{figure} 

   \begin{figure}[ht]   
	\centering
	\includegraphics [width=\columnwidth]{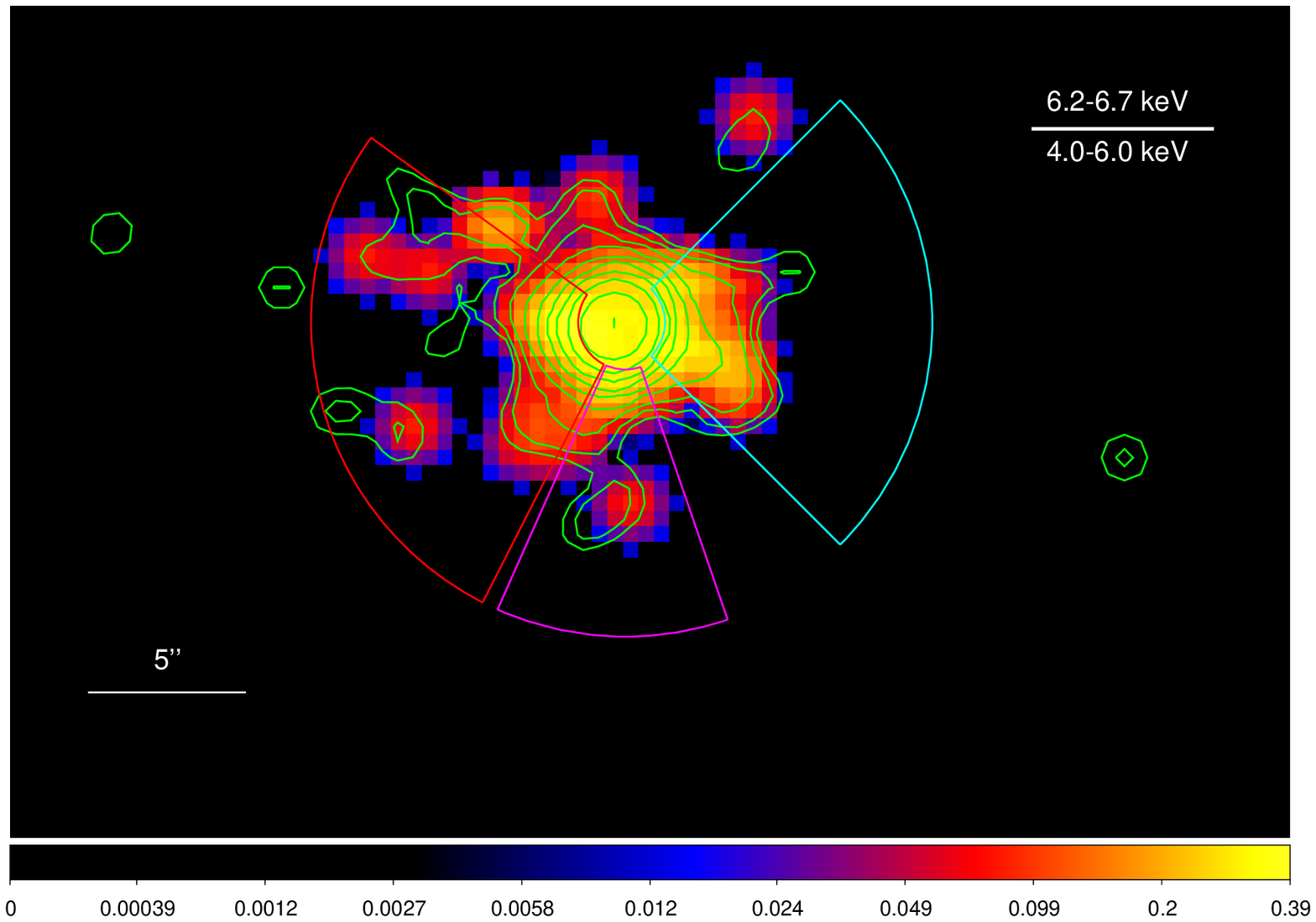}
	\caption{Ratio between the Fe K$\alpha$ and continuum image, with contours extracted from the 6.2--6.7 keV image. The Fe K$\alpha$ EW excess is consistent with the extended structure of Fe K$\alpha$ emitting material. The cones in three different colors ranging from 1.5 arcsec to 10 arcsec in radius show position angles with counts excess. The regions with excess of Fe K$\alpha$ EW appear clumpy and contained in the three cones. We adopted the native ACIS instrumental pixel (0.492'') and smoothed the image with a Gaussian kernel of 3 pixels.}
\end{figure} 

  \section{Results} 
      \subsection{Imaging}

The central 30 arcsec $\times$ 30 arcsec region of NGC 4388 is shown in Figure~1. The images were extracted from the merged event file in energy between 4.0--6.0 keV (Figure 1b) and 6.2-6.7 keV (Figure 1c). The first band was chosen to image the circumnuclear material responsible for Compton reflection continuum excluding any strong line contribution while the other band was chosen to map the neutral Fe K$\alpha$ emission. To quantify the significance of the extended components in these energy bands, we then simulated the ACIS-S PSF contribution by using the simulation tool {\em ChaRT} \footnote{http://cxc.harvard.edu/ciao/PSFs/chart2/} and {\em MARX} \footnote{http://cxc.harvard.edu/ciao/threads/marx/} \citep{2012SPIE.8443E..1AD}. The peak of the PSF emission is normalized to the peak of the source radial profile. Figure 2 shows the radial profiles of both the hard continuum band and Fe K$\alpha$ band. The source emission is extended and well above the expected PSF up to $10$ arcsec ($\sim810$ pc). Furthermore, the extended Fe K$\alpha$ emission, as shown in Figure 1c, is only prominent in certain directions. In order to better identify these extended regions, we provide the number of counts for different azimuthal directions, with angular bins of $18^{\circ}$ (Figure 3).  As can be seen from Figure 3, the position angles showing counts excess in 6.2--6.7 keV are generally consistent with that in 4.0--6.0 keV and can be divided into three parts (hereafter defined as the ``cones''). The east cone (red in Figure 4) is between position angles (measured counter-clockwise from north) $53^{\circ}$ and $152^{\circ}$, the south cone (magenta in Figure 4) between $155^{\circ}$ and $201^{\circ}$, and the west cone (cyan in Figure 4) between $225^{\circ}$ and $315^{\circ}$. These angular extent should not be treated as exact values for such diffuse features. We then divide the 6.2--6.7 keV image by the 4.0--6.0 keV one and smooth the image with a gaussian kernel of 3 pixels (Figure 4). Adopted as a proxy for EW, this ratio map indicates the relative strength of Fe K$\alpha$ emission line with respect to the associated reflection continuum \citep {Fabbiano2017Discovery,Marinucci2017}. The regions with elevated EW values are contained in the cones, corresponding to the same sectors with counts excess shown in Figure 3. 

\begin{table*}
     	\caption{The column density $N_{\rm H}$ of NGC 4388  detected by X-ray instruments for over thirty years}
     	\begin{center}
     		\begin{tabular}{cccc}
     			\hline
     			Instrument& $N_{\rm H}$& Obs.date& Ref. \\
     			&$10^{23}$ \psqcm& & \\
     			\hline
     			SL2-XRT&$2.1^{+2.8}_{-1.4}$&1985 Jul & \citet{1990MNRAS.242..262H} \\
     			ASCA-SIS & $4.2^{+0.6}_{-1.0}$&1993 Jul & \citet{iwasawa1997asca}\\
     			ASCA-GIS& $3.7^{+1.0}_{-0.5}$&1993 Jul &\citet{iwasawa1997asca}\\
     			ASCA & $3.34^{+0.98}_{-0.84}$& 1995 Jun &\citet{1999Forster}\\
     			BeppoSAX-1&$3.80^{+0.2}_{-0.4}$&1999 Jan & \citet{2002ApJ...571..234R}\\
     			BeppoSAX-2&$4.8^{+1.8}_{-0.8}$&2000 Jan &\citet{2002ApJ...571..234R}\\
     			Chandra&$3.5^{+0.4}_{-0.3}$&2001 Jun &\citet{2003MN}\\
     			XMM-1&$2.45^{+0.2}_{-0.21}$&2002 Jul &\citet{2004Beckmann}\\
     			XMM-2&$2.79^{+0.07}_{-0.07}$&2002 Dec &\citet{2004Beckmann}\\
     			Suzaku&$3.39^{+0.15}_{-0.47}$&2005 Dec&\citet{shirai2008detailed} \\
     			NuSTAR$^{\dagger}$ &$6.5^{+0.8}_{-0.8} $ &2013 Dec&\citet{2017ApJ...843...89K}\\
     			NuSTAR &$5.3^{+0.7}_{-0.7}$& 2013 Dec&\citet{2017ApJ...843...89K}\\
     			NICER &$2.67^{+0.02}_{-0.03}$&2017 Dec&\citet{2019Miller}\\
     			NICER$^{\dagger}$ &$2.64^{+0.03}_{-0.03}$ &2017 Dec&\citet{2019Miller}\\
     			\hline   			
     		\end{tabular}
     	\end{center}
     	\tablecomments{$^{\dagger}$ Spectral fit with MYTorus model}.
     \end{table*} 
        
 \begin{table*}
  	\begin{center}
  	\caption{Best-fit parameters for the spectra of nuclear and extended regions}	
  	\begin{tabular}{cccc} 
  		\hline
  		Parameter & Nucleus & Nucleus & Extended \\
  		\hline
  		    & zTBabs$\times$(pow$+$zgauss$+$zgauss)&pcfabs$\times$(pow+zgauss+pexrav+zgauss)& pexrav+zgauss+psf-contribution\\
  		 \hline
  	     $\chi^{2}$/dof	& 1.02& 1.10&1.27\\
  	     $N_{\text{H}}^a$&2.6$^{+0.14}_{-0.14}$&3.6$^{+0.27}_{-0.23}$&\nodata\\
  	     $A_{pow}^b$&4.37$^{+0.3}_{-0.3}$&2.57$^{+0.2}_{-0.2}$ &0.6$^{+0.4}_{-0.4}$ \\
  	     $E_{\rm Fe K_{\alpha}}^c$ & 6.35$^{+0.02}_{-0.02}$&6.35$^{+0.02}_{-0.02}$&6.37$^{+0.04}_{-0.04}$
  	     \\
  	     
  	     $I_{\rm Fe K_{\alpha}}^d$& 7.73$^{+1.03}_{-1.03}$ &8.31$^{+1.15}_{-1.0}$&0.2$^{+0.05}_{-0.05}$\\
  	     
  	     $EW_{\rm Fe K_{\alpha}}^e$&474$^{+71}_{-70}$ &431$^{+77}_{-76}$&1415$^{+330}_{-330}$\\
  	     
  		 $E_{\rm Fe K_{\beta}}^c$&7.01$^{+0.05}_{-0.05}$&7.01$^{+0.07}_{-0.07}$&\nodata\\
  		 
  		 $I_{\rm Fe K_{\beta}}^d$&1.91$^{+0.76}_{-0.76}$&1.68$^{+0.85}_{-0.85}$&\nodata\\
  		 
  		 $EW_{\rm Fe K_{\beta}}^e$&165$^{+68}_{-66}$&118$^{+176}_{-110}$&\nodata\\
  		\hline

  	\end{tabular} 
    
  	\end{center}
 \tablecomments{(a) Absorbing column density \nh, in units of $10^{23}cm^{-2}$. (b) Normalization of the power--law model component in units of $10^{-3}$ photons keV$^{-1}$cm$^{-2}$s$^{-1}$ at 1 keV. (c) Line central energy in keV. (d) Line flux in units of $10^{-5}$ photons cm$^{-2}$ s$^{-1}$. (e) Equivalent widths in eV. See text for the description of spectral models.}

\end{table*}

\subsection{Spectral fitting}\label{spec}  
 
As mentioned in Section 2, the bright nucleus of NGC 4388 was affected by mild photon pile-up (estimated to be $\approx 10$\%). The main consequences of the photon pile-up are a hardening of the spectrum and a reduction in the detected flux. We correct for these effects using the pile-up model of \citet{Davis2001}.  The spectrum of the NGC~4388 nucleus presented in Figure 5a was taken from within 1.5 arcsec radius of the hard X-ray peak (R.A. = $12^{\rm h}25^{\rm m}46^{\rm s}.77$, Dec. =$+12^{\circ}39^{\prime}44^{\prime\prime}.0$ J2000 \citep{2003MN}).  To eliminate contribution from the diffuse, soft X-ray emitting gas, we focus our analysis to the energy band above 3 keV. A simple absorbed \citep[Tuebingen-Boulder ISM absorption model zTBabs,][]{wilms00} power law model is fitted to the 3.0--8.0 keV data (Figure 5b). The photon index of the power-law continuum is assumed to be $\Gamma=1.6$ recently obtained with NuSTAR data in \citet{2017ApJ...843...89K}.  The absorption column density is found to be \nH $\simeq 2.6^{+0.14}_{-0.14}\times 10^{23}$ cm$^{-2}$ which is consistent with previous results \citep{2004Beckmann, 2019Miller,2003MN,2002ApJ...571..234R}.  We obtained a $\chi^2$/dof $ = $187.87/184$ = $1.02, where the dof is the degrees of freedom in fit and no strong residuals throughout the fitted energy band with inclusion of two Gaussian emission lines. 

A strong emission line consistent with iron fluorescent emission (Fe K$\alpha$) is found at an energy of $E_0=6.35^{+0.02}_{-0.02}$ keV (observer's frame) with an EW=$474^{+71}_{-70}$ eV. The line flux is $7.73^{+1.03}_{-1.03}\times10^{-5}$ \phpspsqcm, consistent within tens of percent, with the measurements of the last 20 years. \citep{shirai2008detailed,2004Beckmann,shu2011}. The constant Fe line emission is probably decoupled from the direct emission of the central engine, and is likely produced in distant obscuring torus and beyond \citep{2004Beckmann}. The obscured nucleus has recently been further investigated by \citet{2017ApJ...843...89K}. The 2--10 keV flux observed by {\em NuSTAR} was found to be $\sim8.0\times10^{-12}$ erg cm$^{-2}$ s$^{-1}$ which is in fairly good agreement with our results ($8.21\times10^{-12}$ erg cm$^{-2}$ s$^{-1}$), indicating that the source remained in a similar relatively low flux state during NuSTAR and Chandra observations. Thus, we fit the nuclear spectra with the same {\tt  pexrav} model described in \citet{2017ApJ...843...89K}. In addition to the Compton reflection component {\tt pexrav}, the model also contains an absorbed power law with two Gaussian lines. For the absorption, we adopted a partial covering model {\tt pcfabs} and found a covering factor of 0.95 for the obscuring material. The fitting result is shown in Figure 5c and $\chi^2/dof$ =200.80/182$=$1.10. The EW of Fe K$\alpha$ is $431^{+77}_{-76}$ eV which is consistent with the recent measurement ($368^{+56}_{-53}$ eV) discussed in \citet{2017ApJ...843...89K}.

 There is also another weak line detected at $E_0=7.01^{+0.05}_{-0.05}$ keV with a flux about 5 times lower than the neutral Fe K$\alpha$ line. The line energy implies that it is most likely the Fe K$\beta$ line, but could be blended with Fe XXVI K$\alpha$ at 6.97 keV from a highly ionized medium. The spectrum shows no sign of a further line at 7.1 keV as reported from Chandra data by  
  \citet{2003MN}, in agreement with XMM-Newton data in \citet{2004Beckmann}.  To evaluate the statistical significance of the line centered at 7.01 keV, we run Monte Carlo simulations for our data set following the method discussed in \citet{2010Tombesi} and \citet{2016Walton}.  We adopted the best fit power-law continuum model for nucleus without emission lines to simulate a fake spectrum.  Using the {\tt fakeit} command in XSPEC, 10000 sets of simulated spectra are produced with the same responses, background files, exposure times and energy binning as the ones used for our observations. We fit the fake spectrum using the continuum model and recorded the best fit parameters. A new unresolved Gaussian line was added to the continuum model and the line energy center was free and allowed to vary from 5 keV to 8 keV in 100 steps ({\tt steppar} command in XSPEC). If S is the total simulated spectra and N is the number of data sets for which the fitting results were improved, the estimated statistical significance of this line is 1-$N/S$. For the putative Fe K$\beta$ line at 7.01 keV, out of total 10000 simulations (S), N=207 spectra was found and the confidence level of this line is $1-207/10000=98\%$ (2.33$\sigma$). 
  
  There is no obvious excess indicating the presence of Fe XXV line in the nuclear spectra. We also calculated the statistical significance of this line (1-4856/10000, $<$1$\sigma$). Thus, the Fe XXV line is not further considered in the fitting results (Table 3). We further checked the 6.7--7.1 keV image and find that 93$\%$ counts are confined to the nucleus within two arcseconds, hence there is no significant extended highly ionized component.
  
  \begin{figure}
   \centering
   \includegraphics [width=\columnwidth]{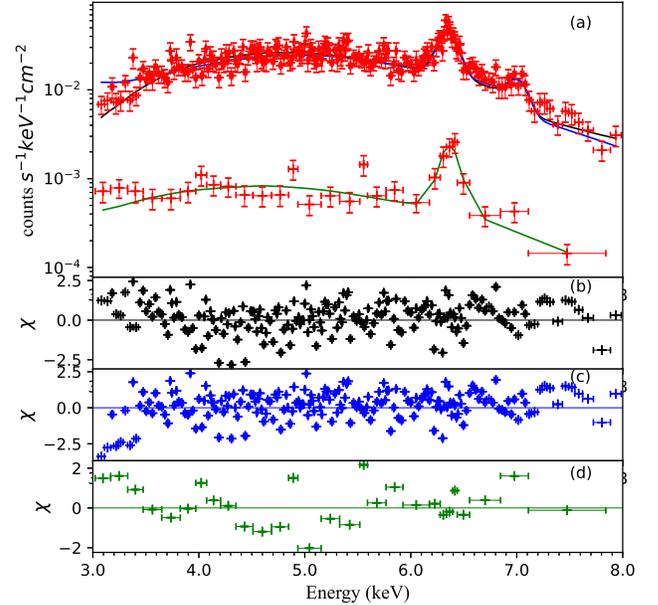}
   \caption{(a) Energy spectra with models fitting for nuclear and extended regions. The black line refers to a simple absorbed power-law model while blue line depicts a {\tt pexrav} model for nuclear region. Green line refers to the extended region model. (b-c) the fit residuals for the simple absorbed power-law and {\tt pexrav} model for nuclear region, respectively. (d) the fit residual for the extended region model.}
   \end{figure}

For the extended regions, in order to take into account the contamination from the PSF wings, we extracted the spectrum from corresponding extended regions using simulated events file of the nucleus, and fitted it with a simple model consisting of an absorbed power-law plus Gaussian line. We create RMF and ARF files (RMF describes the instrument response and ARF the telescope effective area) following the procedure provided by the MARX documentation \citep{2012SPIE.8443E..1AD}. Due to the low number of counts, following \citet{Lanzuisi13}, we binned the spectra using 10 counts per channel and used the Cstat statistic \footnote {Extensive tests on using Cstat with binned data and comparison with $\chi^2$ statistic has been carried out in the Appendix of \citet{Lanzuisi13} }, a modified version of the Cash statistic \citep{1976A&A....52..307C}. The photon index of power-law is again assumed to be 1.6, as measured with the NuSTAR data in \citet{2017ApJ...843...89K} and we obtain the $\chi^2$/dof $=$ 1.25. The normalization of power-law ($5.2^{+2.0}_{-1.5} \times 10^{-5}$ photons keV$^{-1}$cm$^{-2}$s$^{-1}$ at 1 keV) indicates that it is about 1.2 \% of the power-law component from the nucleus. 
      
     \begin{figure}[h]  
        \centering
        \includegraphics [width=\columnwidth]{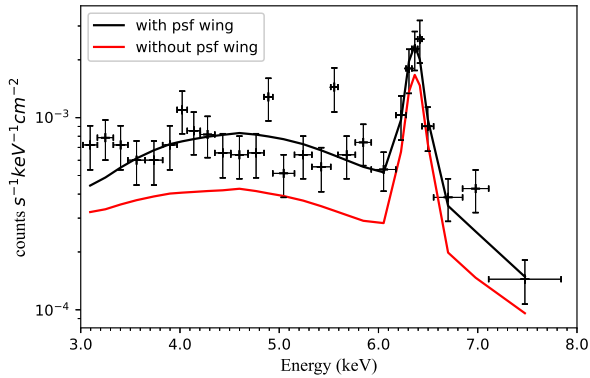}
        \caption{3.0--8.0 keV spectra for extended regions. Black and red line indicate spectral model with and without contribution from nuclear psf wings, respectively.}
        \end{figure} 
     
     In the following steps, the contribution from the PSF wings was included as a fixed model component. The final spectral model for the extended regions also consists of a reflection continuum {\tt pexrav} and a neutral Fe K$\alpha$ line at 6.4 keV. The photon index of {\tt pexrav} was fixed to 1.6 and the cutoff energy $E_{c}$ was fixed to 1000 keV \citep{2017ApJ...843...89K}. We obtained an acceptable fit with $\chi^2$/dof=27.92/22=1.27. The fitting result is shown in Figure 5d. If we remove the psf-contribution from complete spectrum, the rest spectrum with evidence of a strong Fe K$\alpha$ line is shown in Figure 6 (red line). The equivalent width of the Fe K$\alpha$ line (without psf contribution) for extended regions is $1.415^{+0.33}_{-0.33}$ keV. This is different from the inferred EW=$474^{+71}_{-70}$ eV in the nuclear region by a factor $\sim 3$, implying a different geometry or iron abundance of the circumnuclear extended matter in comparison with the nucleus. The observed 6.2--6.7 keV flux is $2.0^{+0.5}_{-0.5}$$\times10^{-6}$ \phpspsqcm and the luminosity is $3.3^{+0.9}_{-0.9}$$\times10^{39}$$ erg/s$. The estimated 3.0--8.0 keV luminosity, excluding the iron line contribution, is 8.61 $\times10^{39}$ \ergps.
     The Fe K$\alpha$ line flux for nucleus is 7.73$^{+1.03}_{-1.03}$$\times10^{-5}$\phpspsqcm and thus the ratio of Fe K$\alpha$ emission line intensity between the extended and nuclear region is 2.6\%. For the hard-band continuum, the flux for extended and nuclear region is 2.36$\times10^{-6}$ \phpspsqcm and 1.69$\times10^{-4} $\phpspsqcm, respectively, corresponding to a ratio of 1.39\%. 
     Thus, the EW of Fe K$\alpha$ line for extended region is about twice than that in nucleus. The best fit parameters for the spectra of nuclear and extended regions are summarized in Table 3.

\begin{table*}
	\caption{Properties of the Seyfert sample with Resolved Fe K$\alpha$ line emission}
	\begin{center}
		\begin{tabular}{ccccccccc}
\hline
$Sources$ & $N_{\rm H}$ & $log(L_{\rm [O IV]})$ &  $log(L_{\rm [Fe K\alpha]tot})$ &$log(L_{\rm [Fe K\alpha]ext})$ &ratio&$EW_{[Fe K\alpha]ext}$& $log(L_{\rm [Fe K\alpha]ext}/EW_{[Fe]ext})$& \\ 
 & $10^{23}cm^{-2}$ & $erg/s$ & $erg/s$ & $erg/s$ & &$keV$& $erg/s$&  \\
 \hline
 NGC 4945 & $22.0^{a}$ & $39.37^{+0.12}_{-0.17}$$^{f}$ & $39.32^{+0.04}_{-0.03}$$^{h}$& $38.18^{+0.05}_{-0.05}$$^{j}$ & 0.07&$1.10^{+0.13}_{-0.12}$$^{j}$&$38.13^{+0.06}_{-0.06}$&\\

 Circinus & $43.0^{a}$ & $40.50^{+0.12}_{-0.17}$$^{f}$ & $40.14^{+0.01}_{-0.01}$$^{h}$&$38.82^{+0.04}_{-0.05}$$^{k}$ & 0.05 & $0.9^{+0.3}_{-0.3}-2.2^{+0.5}_{-0.5}$$^{k}$&$38.65^{+0.07}_{-0.07}$$^{k^{*}}$& \\

 NGC 1068 & \textgreater $100^{b}$& $41.78^{+0.07}_{-0.08}$$^{f}$&$40.23^{+0.03}_{-0.04}$$^{h}$& $39.62^{+0.08}_{-0.09}$$^{l}$ &0.25&$1.36^{+0.46}_{-0.34}$&$39.51^{+0.17}_{-0.17}$&\\
 NGC 4388 & $2-7^{c}$ & $41.61^{+0.01}_{-0.01}$$^{f}$ & $41.00^{+0.11}_{-0.12}$$^{h}$ &$39.52^{+0.1}_{-0.13}$&0.03 &$1.415^{+0.33}_{-0.33}$&$39.37^{+0.15}_{-0.15}$& \\
 NGC 5643 & \textgreater $50^{d}$ & $40.46^{+0.03}_{-0.04}$$^{f}$ & $39.70^{+0.09}_{-0.10}$$^{h}$ & $39.28^{+0.07}_{-0.09}$$^{m}$ & 0.38&$1.6^{+0.5}_{-0.5}$&$39.07^{+0.16}_{-0.16}$\\
 ESO-428-G014 & \textgreater $100^{e}$ & $41.27^{+0.12}_{-0.17}$$^{g}$ &$39.75^{+0.03}_{-0.05}$$^{i}$&$39.00^{+0.1}_{-0.12}$$^{i}$& 0.18&$2.1^{+0.52}_{-0.52}$ &$38.70^{+0.15}_{-0.15}$\\
 \hline			
			
\end{tabular}
	\end{center}
	\tablecomments{Column 1 : the names of galaxies discovered so far to have the extended Fe K$\alpha$ line emission.  Column 2: the value of the absorbing column density; references (a) \citet{2008Della} (b) \citet{2011Burlon} (c) Table 2 (d) \citet{2015Annuar} (e) \citet{Fabbiano2017Discovery}. Column 3: [OIV] luminosity in erg/s; references (f) \citet{2009Diamond-Stanic}  (g) \citet{2016Asmus} A 20\% calibration uncertainty is adopted as 1 $\sigma$ error. Column 4: the total Fe K$\alpha$ line luminosity in erg/s measured by XMM-Newton; references (h) \citet{2010Liu} (i) \citet{2018Fabbiano}. Column 5: the extended Fe K$\alpha$ line luminosity in erg/s; references (j) this work (k) For Circinus galaxy, the whole extended Fe K$\alpha$ region is divided into six smaller parts in \citet {2013Marinucci}. Thus, the total luminosity of the extended Fe K$\alpha$ is the sum, i.e. $L_{\rm [Fe K\alpha]ext}$$=$$\sum_{j=1}^N L_{\rm [Fe K\alpha]extj}$ ($N=6$) (l) \citet{2015Bauer} (m) \citet{Fabbiano2018} (i) \citet{2018Fabbiano}. Column 6: the ratio of the extended and total Fe K$\alpha$ line luminosity. Column 7: the EW of extended region. (k) For Circinus galaxy, \citet {2013Marinucci} gives the EWs for six smaller extended regions, respectively. Here, we list the range of values. Column 8: 6.4 keV monochromatic luminosity (reflective spectrum) for the extended region. ($k^{*}$) For Circinus galaxy, the total reflective spectrum of extended region is the sum of six smaller regions discussed in \cite {2013Marinucci} i.e. $L_{\rm [Fe K\alpha]ext/EW_{[Fe]ext}}$$=$$\sum_{j=1}^N L_{\rm [Fe K\alpha]extj/EW_{[Fe]extj}}$ ($N=6$). The quoted error is the 90$\%$ confidence interval.} 
\end{table*} 

  \begin{table*}
	\caption{Linear regressions for [OIV] to Fe K$\alpha$ luminosity and 6.4 keV monochromatic luminosity.   }
	\begin{center}
		\begin{tabular}{ccccccc}
			\hline
			Correlation & $a$ & $b$&$a_{best}$ &$b_{best}$&$R^{2}_{best}$&Reference \\
			\hline
			log($L_{\rm OIV}$)-log$(L_{\rm [Fe K\alpha]tot})$& $0.92^{+0.08}_{-0.08}$&$3.02^{+0.36}_{-0.36}$& \nodata &\nodata&\nodata&1  \\   
			&$0.73^{+0.12}_{-0.12}$& $10.24^{+0.45}_{-0.45}$& \nodata&\nodata&\nodata&2 \\
			&  $0.48^{+0.09}_{-0.09}$ & $20.6^{+4.05}_{-4.05}$ &0.51 &19.34 &0.67&3 \\
			
			\hline
			log($L_{\rm OIV}$)-log$(L_{\rm [Fe K\alpha]ext})$&$0.53^{+0.09}_{-0.09}$	 &$17.35^{+4.11}_{-4.11}$ &0.59&14.89 &0.93&3 \\
			\hline	
			log($L_{\rm OIV}$)-log($L_{Fe}/EW_{Fe})_{ext}$	&$0.49^{+0.12}_{-0.12}$&$18.98^{+5.1}_{-5.1}$&0.54 &16.8&0.92&3\\
			\hline   			
		\end{tabular}
	\end{center}
	\tablecomments{$a$ and $b$ represent the mean regression coefficients (slope) and regression constants (intercept) in the fitting results, respectively. $a_{best}$ and $b_{best}$ correspond to the best fit parameters (slope and intercept) with the maximum correlation coefficients. The uncertainties for a and b correspond to 3$\sigma$ errors. References: (1) Compton-Thin Seyfert 2 sample in \citet{2010Liu}; (2) Compton-Thick Seyfert 2 sample in \citet{2010Liu}; (3) this work.}
\end{table*}

\section{Discussion}  

\subsection{Origin of the Extended Iron K$\alpha$ Line Emission with large EW}

It is not a total surprise to find extended Fe K$\alpha$ line beyond the sub-pc scale torus region in an AGN, as bright Fe K$\alpha$ line in the molecular clouds in our Galactic center has been discovered by Koyama et al. (1996) for over two decades. There is strong evidence that the origin of the spatially resolved, variable Fe K$\alpha$ emission from multiple molecular clouds is a reflection of the primary X-ray flare from Sgr A* in the past \citep{Nakajima09,Dogiel09,Ponti10}.

In the last few years, much evidence for the presence of extended Fe K$\alpha$ emission has been found in several nearby Compton thick AGN, such as NGC 1068\citep{2015Bauer}, Circinus \citep{2013Marinucci}, NGC 4945 \citep{Marinucci2017}, ESO 428-G014 \citep{Fabbiano2017Discovery} and NGC 5643 \citep{Fabbiano2018}. These findings show that the reflector may extend on scales ranging from tens to thousands of parsecs. 

As shown in Figures~1 and 2, such a circumnuclear reflecting structure for NGC 4388 is spatially resolved by Chandra observation and extends to scale of $\sim$0.8 kpc. A major difference between NGC 4388 and the former galaxies is that the nucleus of NGC 4388 is a Compton-thin AGN. We have carefully considered the contamination in the extended regions from the nuclear radiation, taking into account the contribution from wings of the PSF. A significant difference in the EWs of Fe K$\alpha$ line is found when nuclear and extended regions are compared (a factor $\approx$ 3.0; see Table 3). Such spatial variations in Fe K$\alpha$ EW can be ascribed to iron abundance \citep{1997Matt}, the angle $\theta_i$ between polar direction and the line of sight \citep{1991A&A...247...25M,1991George}, and the column density of the illuminated material \citep{2012MNRAS.423.3360Y}, as discussed in several works using models with a toroidal \citep{2010MNRAS.401..411Y} or a slab geometry \citep{1997Matt}.

We first examined if the variation in Fe K$\alpha$ EW for nuclear and extended regions is due to anomaly in iron abundance. As previously discussed in \citet{2013Marinucci}, achieving a difference of a factor 2 in EW of Fe K$\alpha$ line requires a difference of 2-5 in iron abundance \citep{1997Matt}. Fitting with the {\tt pexmon} model in XSPEC, we inferred comparable Fe K$\alpha$ abundances A(Fe)$_{\rm nucleus}=0.36^{+0.08}_{-0.15}$ and A(Fe)$_ {\rm extend}=0.45^{+0.72}_{-0.18}$ (with respect to solar value) for the nuclear and extended regions, respectively. Typically, a Type Ia SN produces 5-10 times more Fe than Type II SN. Thus, the enrichment due to iron is mainly attributed to SN~Ia on time-scale of 1 Gyr (late SNIa) or hundreds of million years (early SNIa) \citep{2006ApJ...648..230L}. Such an iron enrichment on these time-scales unlikely remain confined in a region of hundreds of parsecs but rather across the galactic disk. Thus, the difference of EW for nuclear and extended regions due to variations of metallicity is not favored. 

\begin{figure}[h]  
	\centering
	\includegraphics[width=\columnwidth]{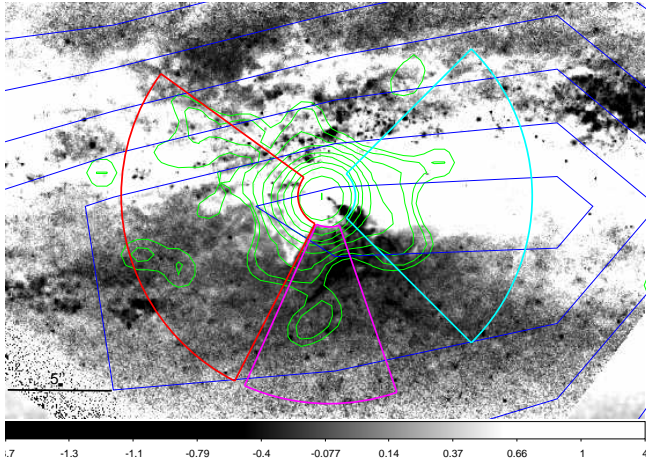}
	\caption{HST V-H color map (grey scale) effectively uncovering dust features (white area) for NGC 4388. The cones in three different colors ranging from 1.5 arcsec to 10 arcsec in radius show position angles with count excess. Dusty regions appear white in the red and cyan cones. The blue contours on behalf of CO(1-0) morphology. The Fe K$\alpha$ extended regions (green contours) are located at the peak CO(1-0) emission and the orientation of north-east and west Fe K$\alpha$ regions are consistent with the galactic disk. }
\end{figure} 

Following \citet{2013Marinucci}, we then tested whether the difference in EW arise from different column densities or inclination angles with respect to the line of sight, using MYTORUS model\footnote{http://mytorus.com/mytorus-instructions.html} \citep{Murphy2009,2010MNRAS.401..411Y} 
to fit the spectra from nuclear and extended regions. The circumnuclear obscuring matter in this model is arranged uniformly in a toroidal geometry with a fixed half-opening angle of 60 degree. Generally, MYTORUS model includes three components: zeroth-orde continuum(MYTZ) with the photons neither absorbed nor scattered, the scattered continuum (MYTS) and fluorescent emission-line spectrum (MYTL). Furthermore, it treats both line and continuum components self-consistently. For all three components (MYTZ, MYTS, MYTL), we link the neutral hydrogen column density $N_H,_{global}$, intrinsic power-law slope $\Gamma$, inclination angle $\theta_{\rm inc}$ and normalization of the  MYTZ,  MYTS, and  MYTL components together. The best fit parameters are summarized as follows:  N$_H,_{global}= 4.2^{+8.3}_{-1.5}\times10^{23}$ \psqcm, $\theta_i = 71.9^{+8.7}_{-10}$ degree for the nucleus ($\chi^{2}_{\nu} = 0.89$) and N$_H,_{global} = 8.9^{+12.0}_{-5.0} \times 10^{23}$ \psqcm, $\theta_i<61.2$ degree for the extended regions  ($\chi^{2}_{\nu} = 0.98$). N$_H,_{global}$ is termed as the equatorial column density angle-averaged over all directions. Thus, it is generally assumed  to be the same as the line-of-sight column density (N$_H,_{Los}$). We caution that N$_H,_{Los}$ can be estimated using N$_H,_{global}$ and $\theta_{inc}$ as shown in \cite{Murphy2009} and in this  paper N$_H,_{Los}$ is simplified to \nh. The N$_H$ and $\theta_i$ are not well constrained by the $<$10 keV Chandra data alone. Nevertheless, we show that the combination of column density and inclination angle could account for the observed spectra as well as the different EWs. This requires presence of some dense gas with column N$_H\sim 10^{23}-10^{24}$ cm$^{-2}$ at the resolved spatial scale in the galactic disk, which is identified next.

It is worth noting that the best fit value of hydrogen column density N$_H$ with MYtorus model is consistent with the reported NuSTAR measurements \citep{2017ApJ...843...89K} within error bars. We have summarized previously published column density measured by various X-ray missions in Table 2. It is noted that the column density estimated by MYtorus model \citep{Murphy2009} is higher than others, which could be due to the adopted more complex geometry and physics other than a simple absorption.  As the upper bound of our measured N$_H$ reaches $10^{24}$ cm$^{-2}$, it is plausible that the measured column density of NGC 4388, generally classified as Compton thin, is statistically consistent with being Compton thick.

\begin{figure}[h]  
	\centering
	\includegraphics[width=\columnwidth]{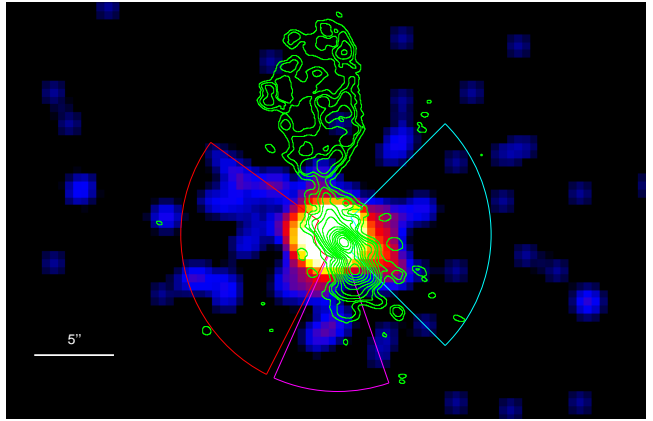}
	\caption{The 6.2--6.7 keV smoothed image with VLA 3.5 cm radio contours \citep{1998ApJ...502..199F} overlaid. The cones in three different colors ranging from 1.5 arcsec to 10 arcsec in radius that show counts excess position angles. The radio jet may squeeze ambient gas and produce high density clumpy cloud to be the reflector to account for the remaining extended Fe K$\alpha$ emission (contained in the south magenta cone and part of the west cyan cone).}
\end{figure}  

Figure 7 shows $V-H$ color map obtained with $HST$ images from Hubble Legacy Archive\footnote{\url{https://hla.stsci.edu/}} which effectively uncovers dust features \citep{2004ASSL..319..213M}. It indicates that the north-east and west extended regions contain rich dust lanes (white area in Figure~7 with inverted color). Furthermore, the $^{12}$ CO(1-0) map observed by IRAM-30 m telescope is also overlaid in Figure 7 (blue contours) \citep{2012A&A...545A..75P}. Although the spatial resolution of CO map is much lower than Chandra, it shows that the extended Fe K$\alpha$ emission line regions (green contours) located at the peak of CO(1-0) molecular gas distribution, which also shows that the orientation of extended north-east and west regions are along the disk of NGC~4388.  \citet{2012A&A...545A..75P} converted the $^{12}$ CO(1-0)  integrated intensity into molecular gas masses (M$_{H2}=7.13^{+0.6}_{-0.6} \times 10^{8}$ M$_\odot$) using the standard Galactic conversion factor $X_{\rm CO}$ = $N_{H_2}/I_{CO} = 2 \times 10^{20} cm^{-2}$ [K km s$^{-1}$]$^{-1}$ \citep{1996Strong,2001Dame}. We further estimated the line-of-sight H$_2$ column density N(H$_2$) $\approx 1.8 \times 10^{21} cm^{-2}$, orders of magnitude lower than the neutral hydrogen column density N$_H$ towards the AGN derived from X-ray observations ($\textgreater 10^{23} cm^{-2}$).   However, this column density estimated from CO(1-0) is averaged over a large area ($117^{\prime\prime}\times 42^{\prime\prime} $) \citep {2012A&A...545A..75P} and should be taken as a lower limit of the true column density. At the peak of the CO(1-0) emission, the value of the column density N(H$_2$) could be much higher. Overall it is consistent with the N (H$_2$) found in other Seyfert galaxies \citep{2020Feruglio,Alonso-Herrero2018,2018Izumi,Fabbiano2018}. In addition, N(H$_2$) is derived for a given conversion factor, which could be different in the nuclei of galaxies other than our Galaxy \citep{2018Wada}. 

We further note that the remaining extended regions (contained in the south and part of the west region) could trace the edges of a known radio outflow. Figure~8 shows the 3.5 cm radio contours \citep{1998ApJ...502..199F} superimposed on the Fe K$\alpha$ image. The extended Fe K$\alpha$ signal suggests that the radio outflow in kpc scale may have compressed the ambient interstellar gas in the galaxy to dense clumps and effectively enhanced reflection and line emission. This is similar to the case in Hydra A \citep{Nulsen_2002} for which the expanding radio lobes compress the surrounding cooler gas.

Last but not least, in contrast to the results in \citet{2003MN}, we find a relatively smaller extent of Fe K$\alpha$ emission ($\sim$10 arcsec or $\sim 0.8$ kpc). Due to the low signal to noise ratio of their Fe K$\alpha$ emission data, \citet{2003MN} adopted 2-arcsec binning for Fe K$\alpha$ image and the low surface brightness extended emission to a few kpc is only 0.05 percent of the nuclear brightness. Such weak signals are less significant in our observation (ObsId 12291) and the deeper merged data, and the very extended regions show surface brightness comparable with background. Hence, we only focus on the most significant extended Fe K$\alpha$ emission line regions in this work.

\begin{figure*}[!htb]
	\subfigure[]{\includegraphics[width=\columnwidth]{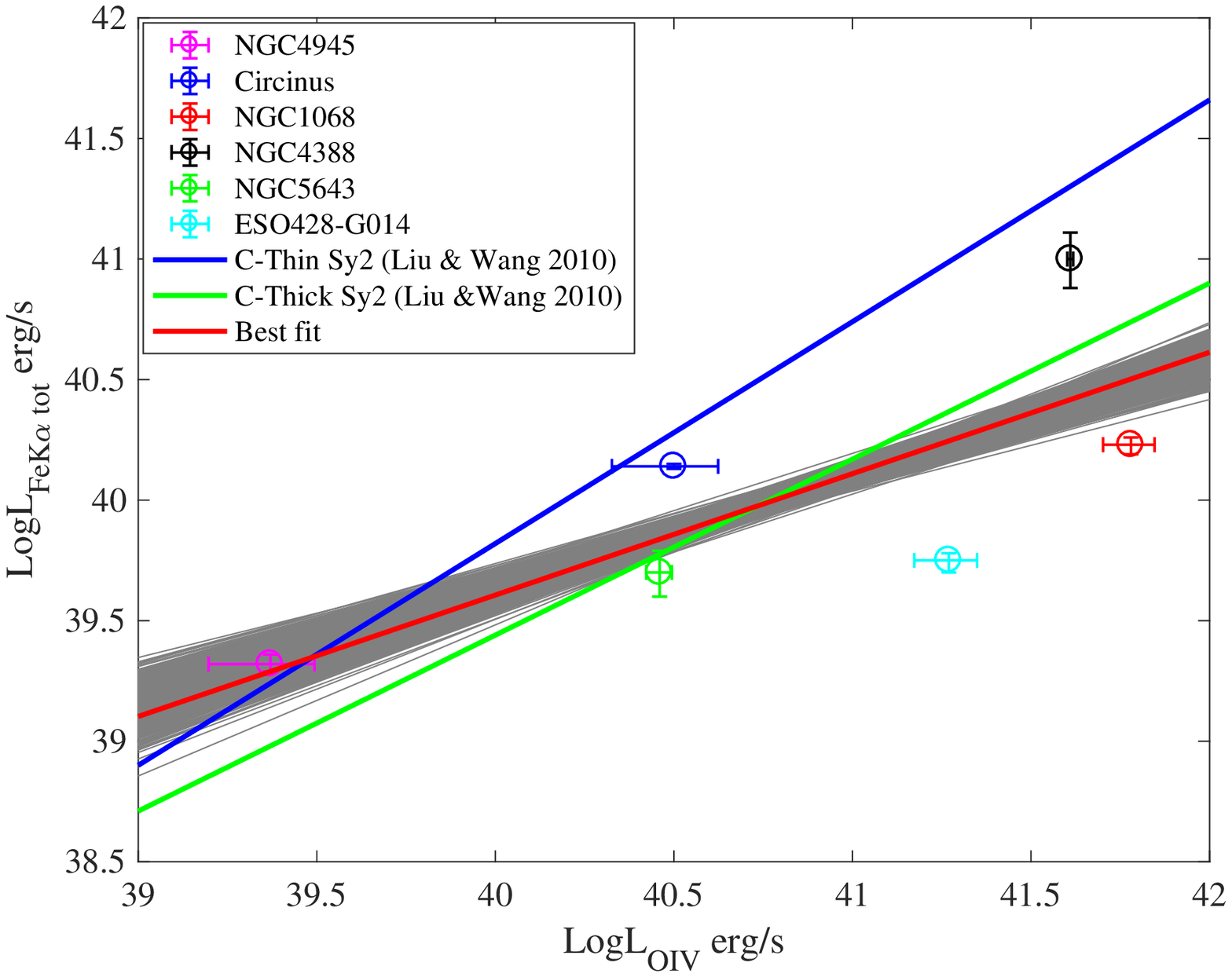}}
	\subfigure[]{\includegraphics[width=\columnwidth]{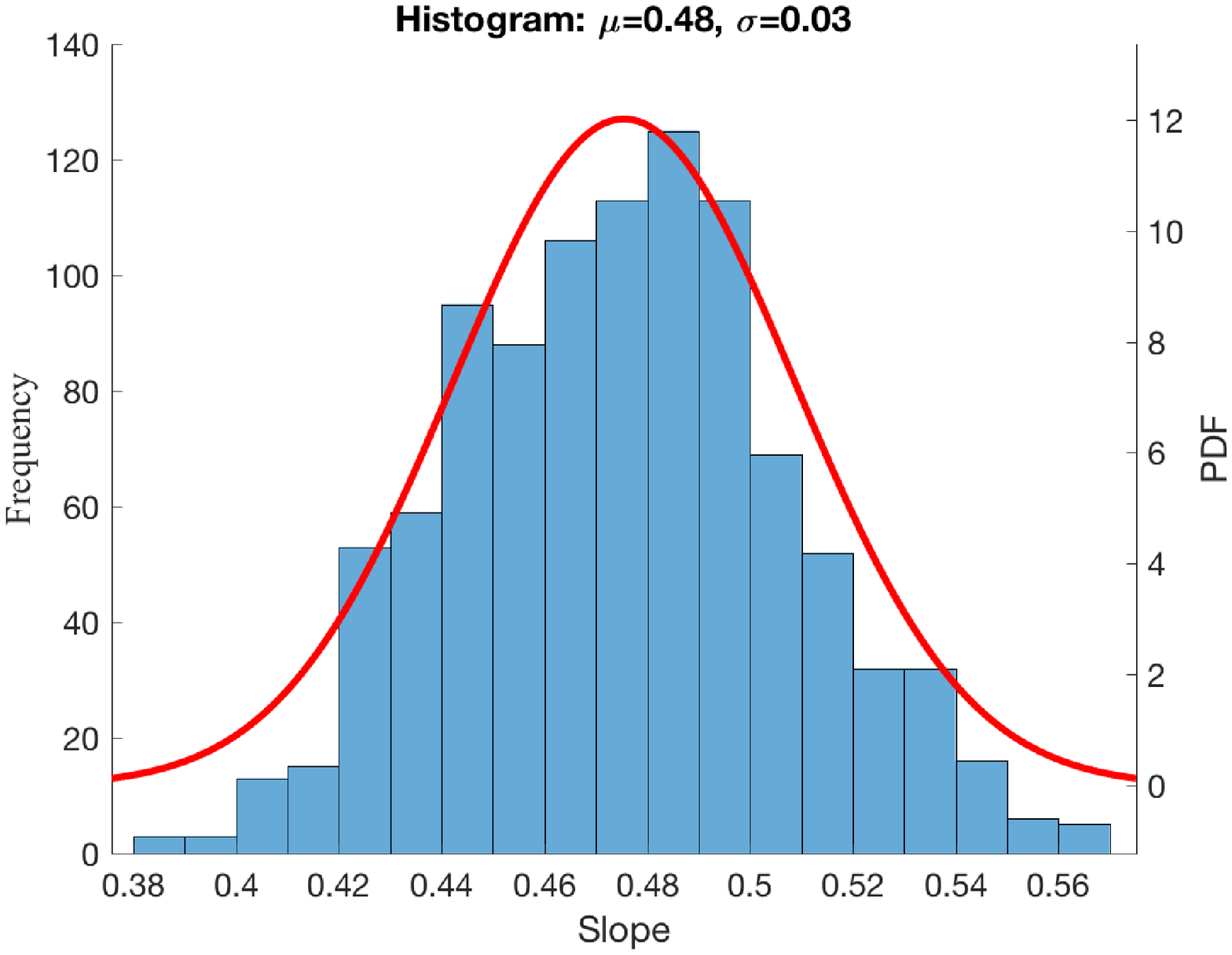}} \\
	\subfigure[]{\includegraphics[width=\columnwidth]{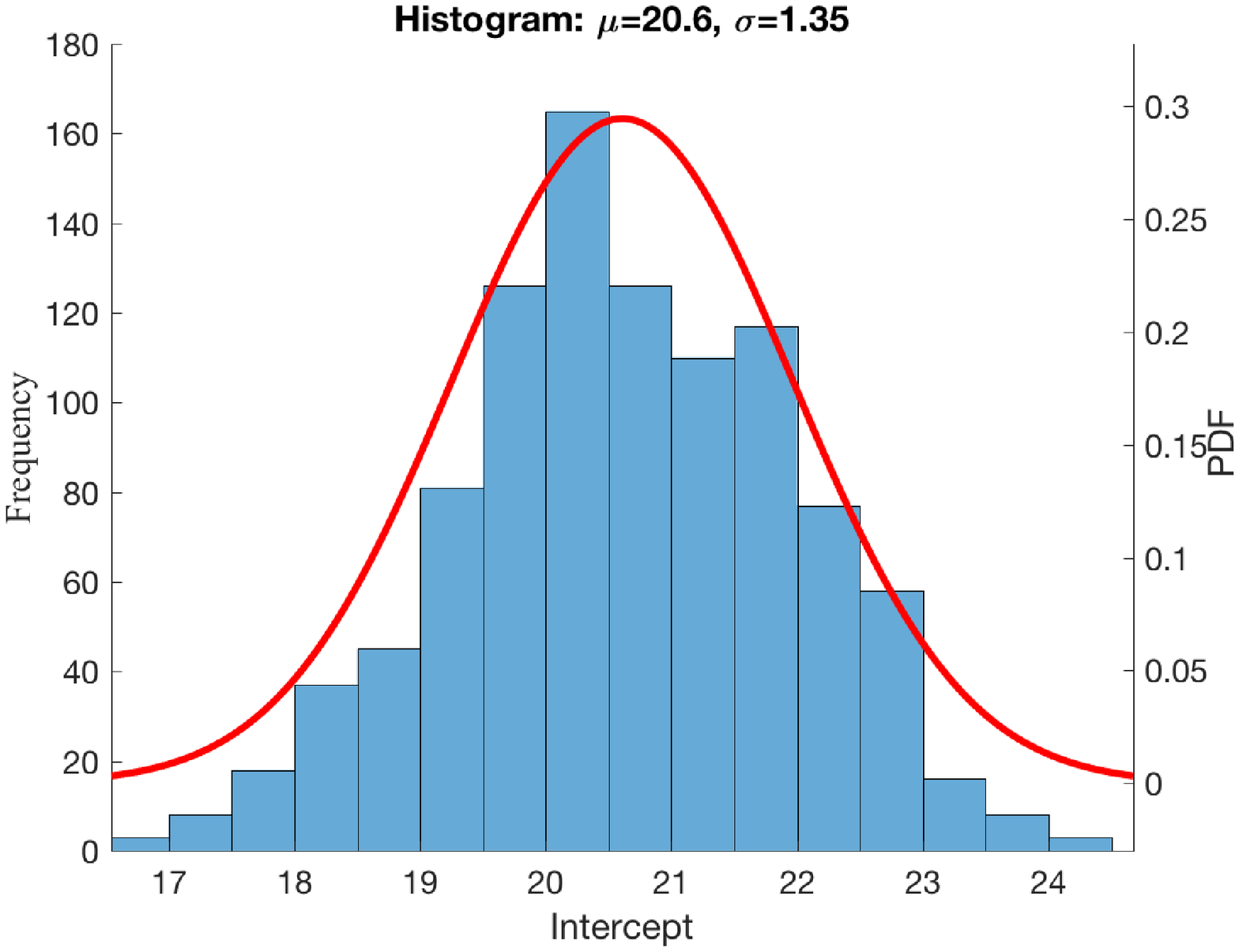}} 
	\subfigure[]{\includegraphics[width=\columnwidth]{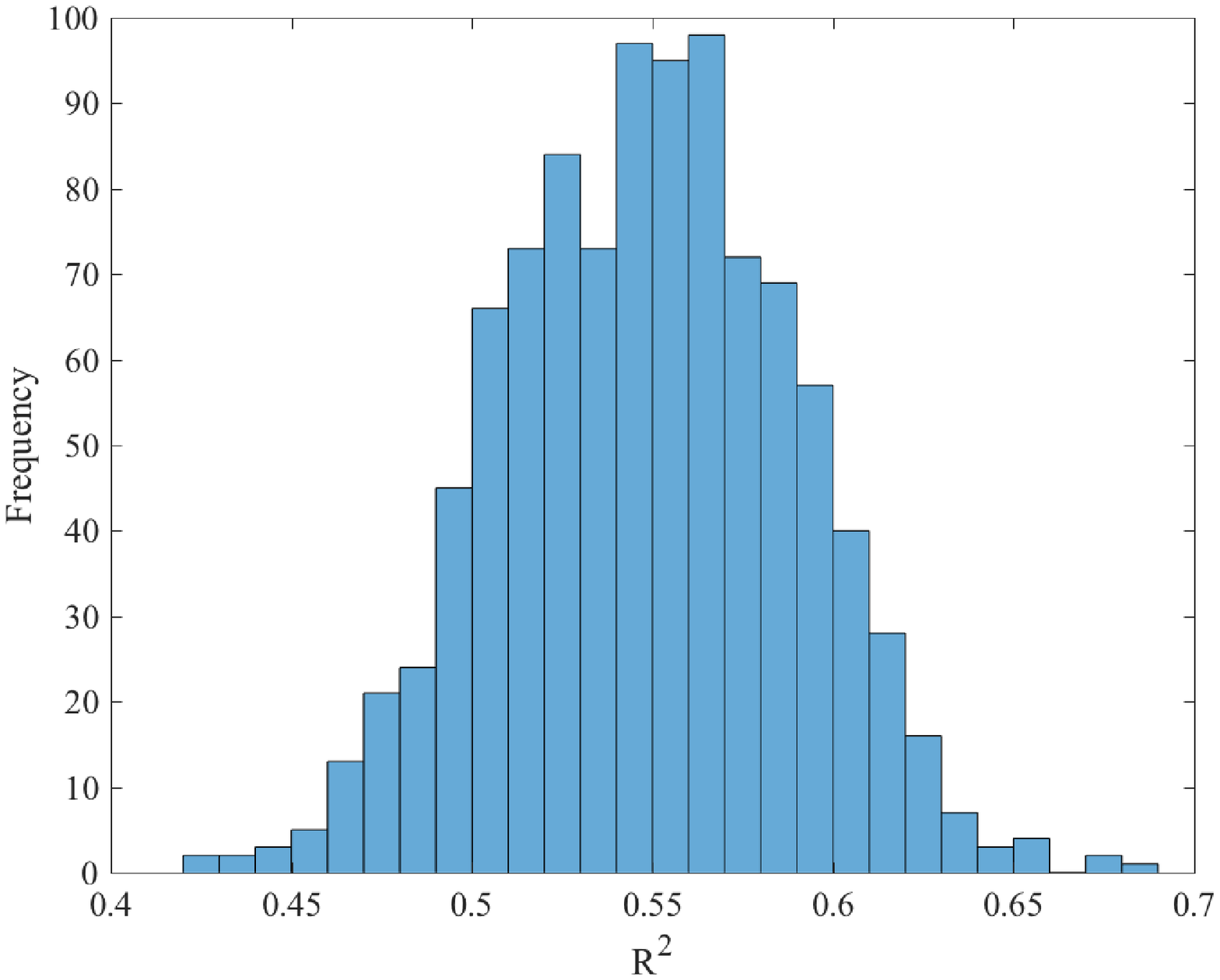}} \\

	\caption{(a) The total Fe K$\alpha$ line luminosity versus log($L_{\rm OIV}$). The blue and green lines are the best fit lines for Compton-Thin Sy2s and Compton -Thick Sy2s in \citet{2010Liu}.  Considering the overall impact of uncertainty for each data on the fitting result, we adopt  Monte Carlo simulations. Assuming the error for each data follows Gaussian distribution, 1000 fitted lines are shown in gray with different combination of six points. The red line represents the best fitted result. (b) the slope distribution from our Monte Carlo analysis.
	We fitted it by a Gaussian model (red line) and obtain the parameters ($\mu$ and $\sigma$) of probability density function(PDF) (c) the intercept distribution from our Monte Carlo analysis.
	(d) the linear regression coefficient distribution from our Monte Carlo analysis.}
	 
	\vspace{0.02in}
\end{figure*}

\begin{figure*}[!htb]
	\subfigure[]{\includegraphics[width=\columnwidth]{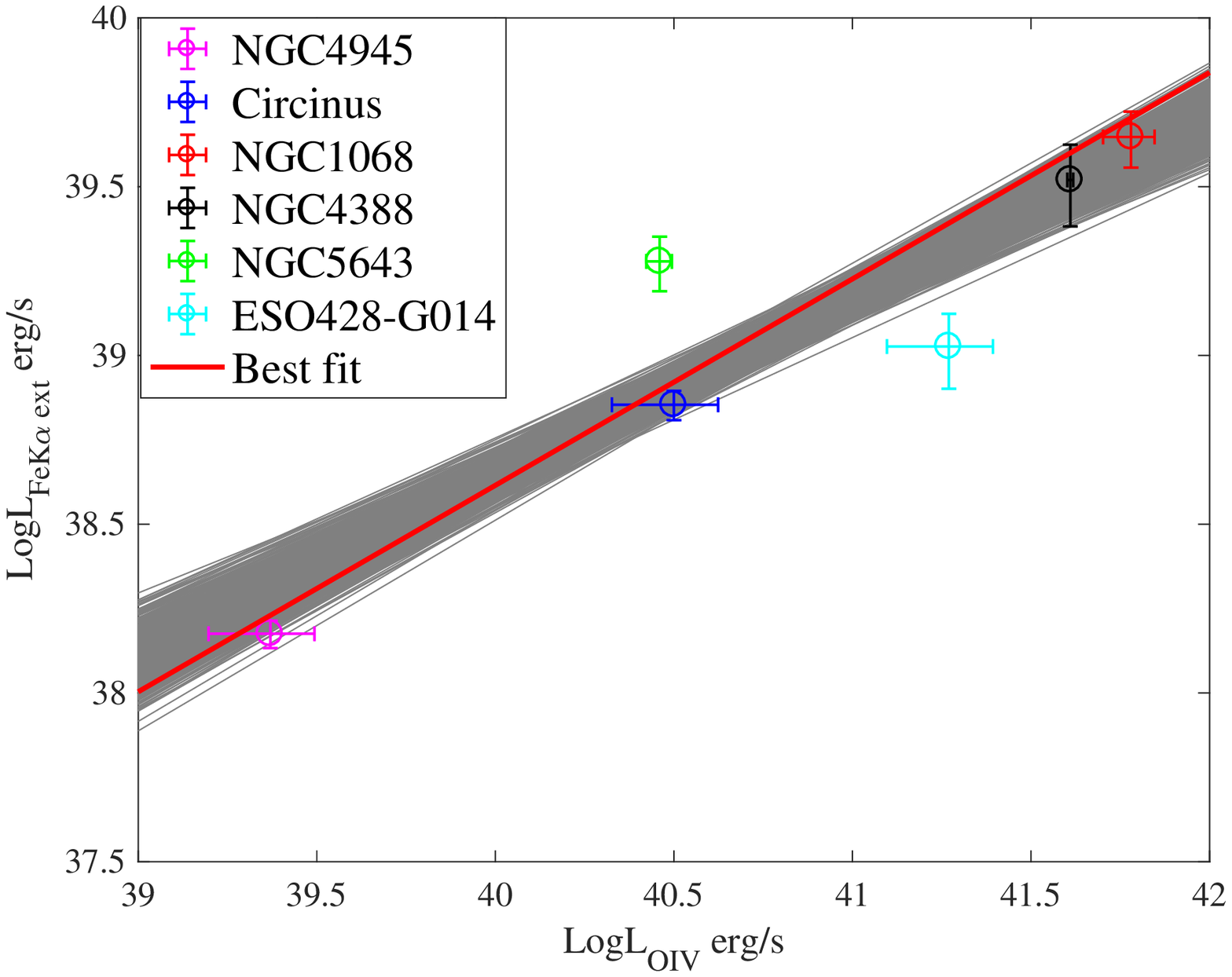}}
	\subfigure[]{\includegraphics[width=\columnwidth]{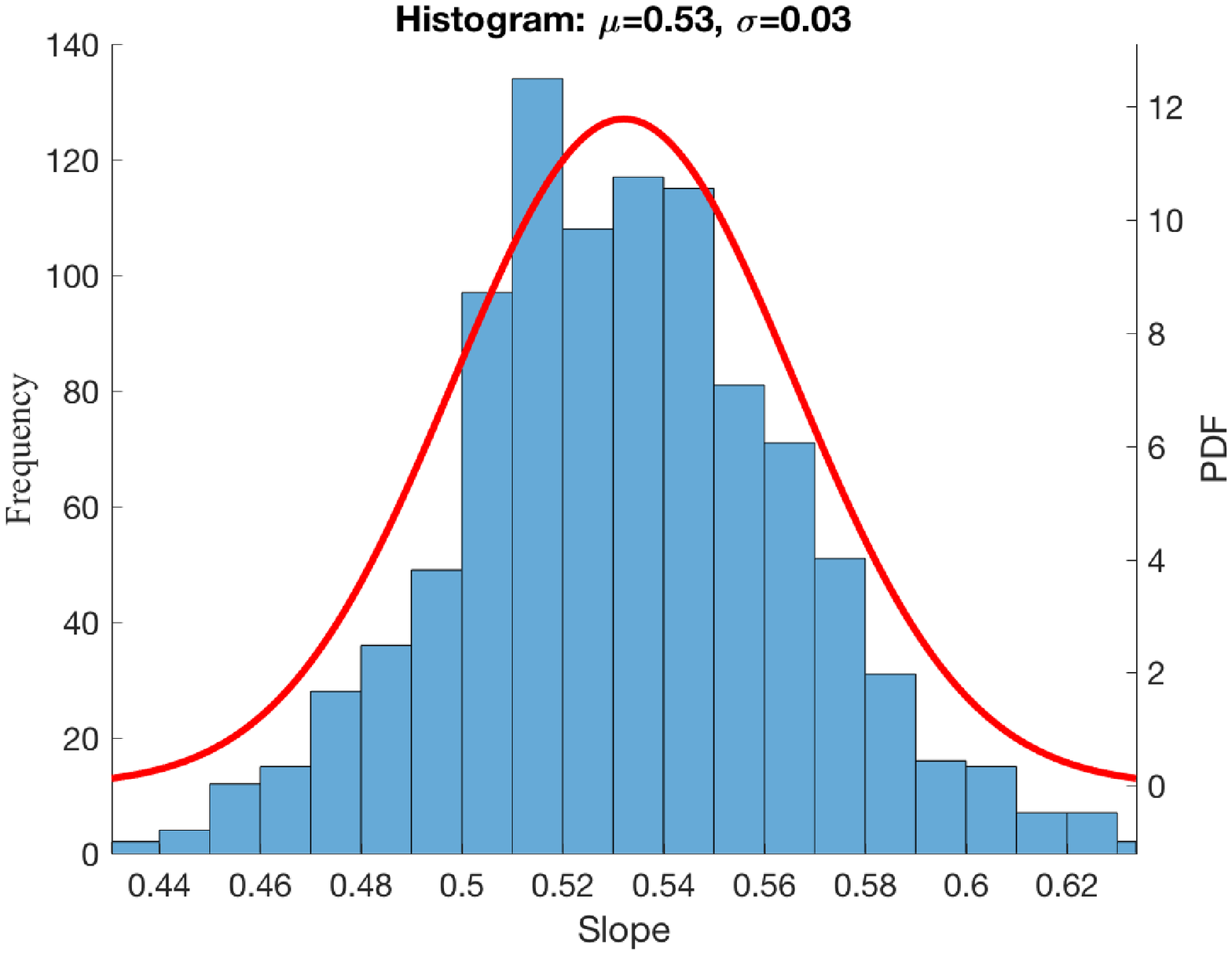}} \\
	\subfigure[]{\includegraphics[width=\columnwidth]{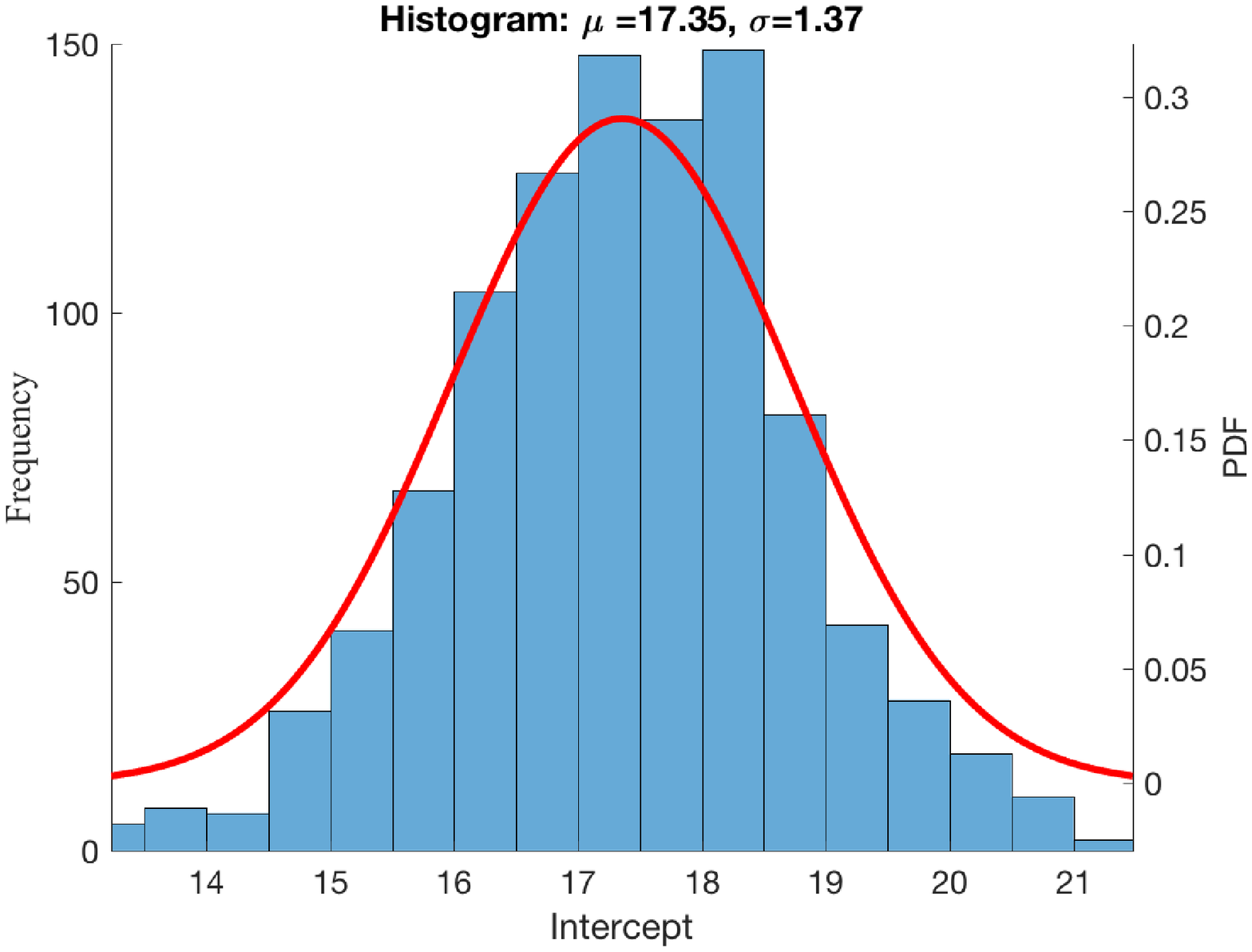}} 
	\subfigure[]{\includegraphics[width=\columnwidth]{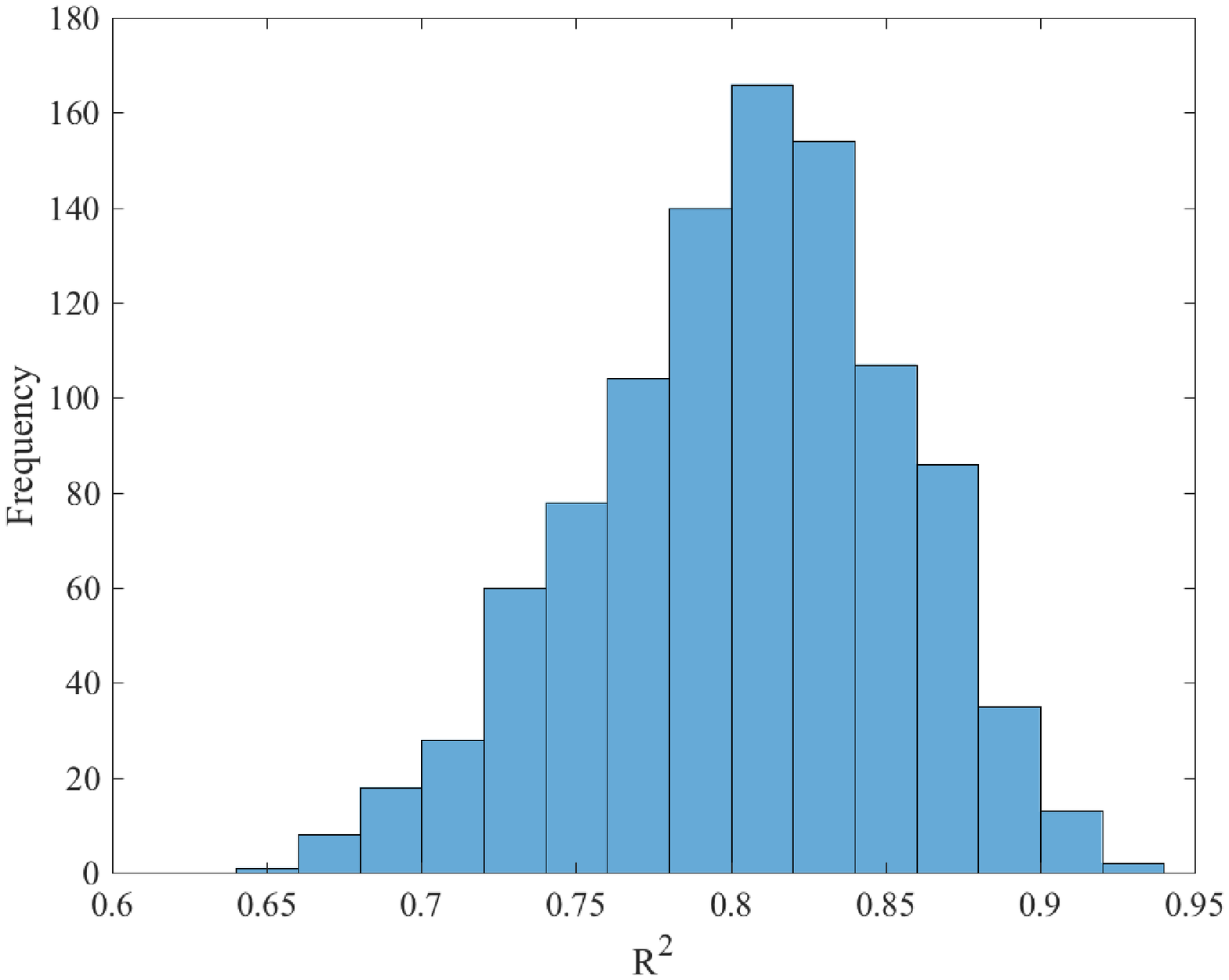}} \\

	\caption{(a) The extended Fe K$\alpha$ line luminosity versus log($L_{\rm OIV}$). Similarly, we run 1000 Monte Carlo simulations. The gray region represent the fitting results. (b) the corresponding linear slope distribution. Red line represents the fitted Gaussian model. (c) intercept distribution. (d) linear regression coefficient distribution.}
	\vspace{0.02in}
\end{figure*} 

\begin{figure*}[!htb]
	\subfigure[]{\includegraphics[width=\columnwidth]{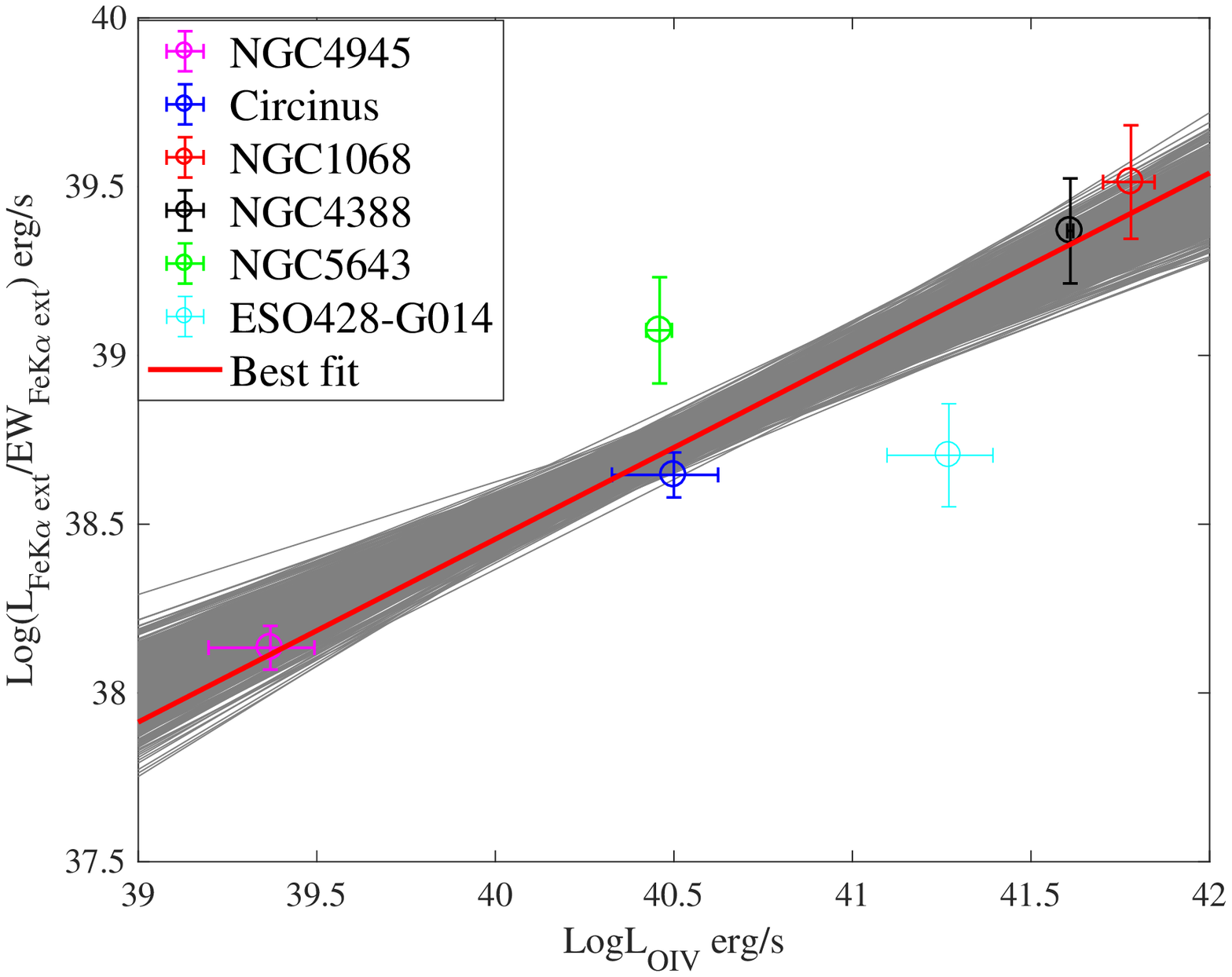}}
	\subfigure[]{\includegraphics[width=\columnwidth]{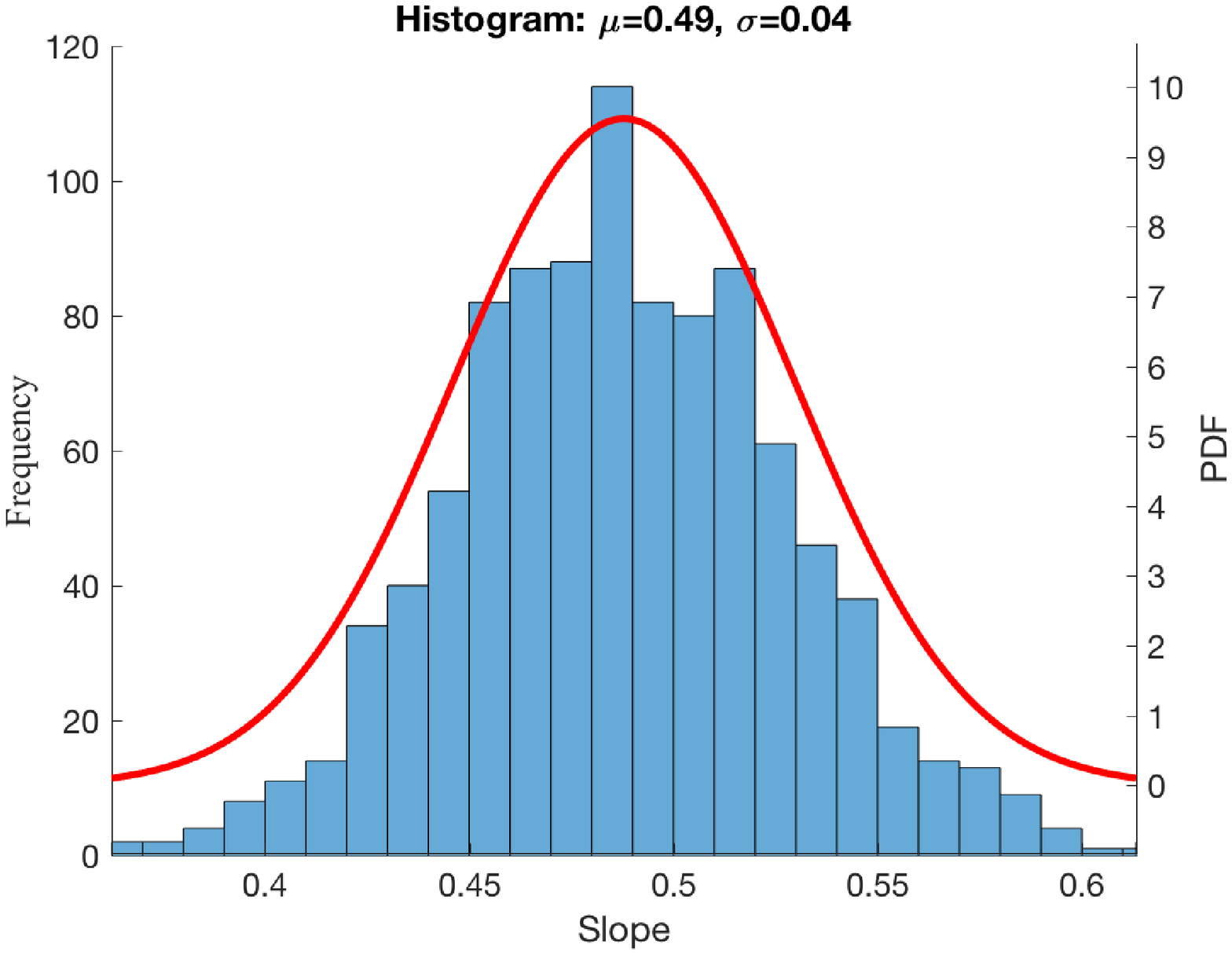}} \\
	\subfigure[]{\includegraphics[width=\columnwidth]{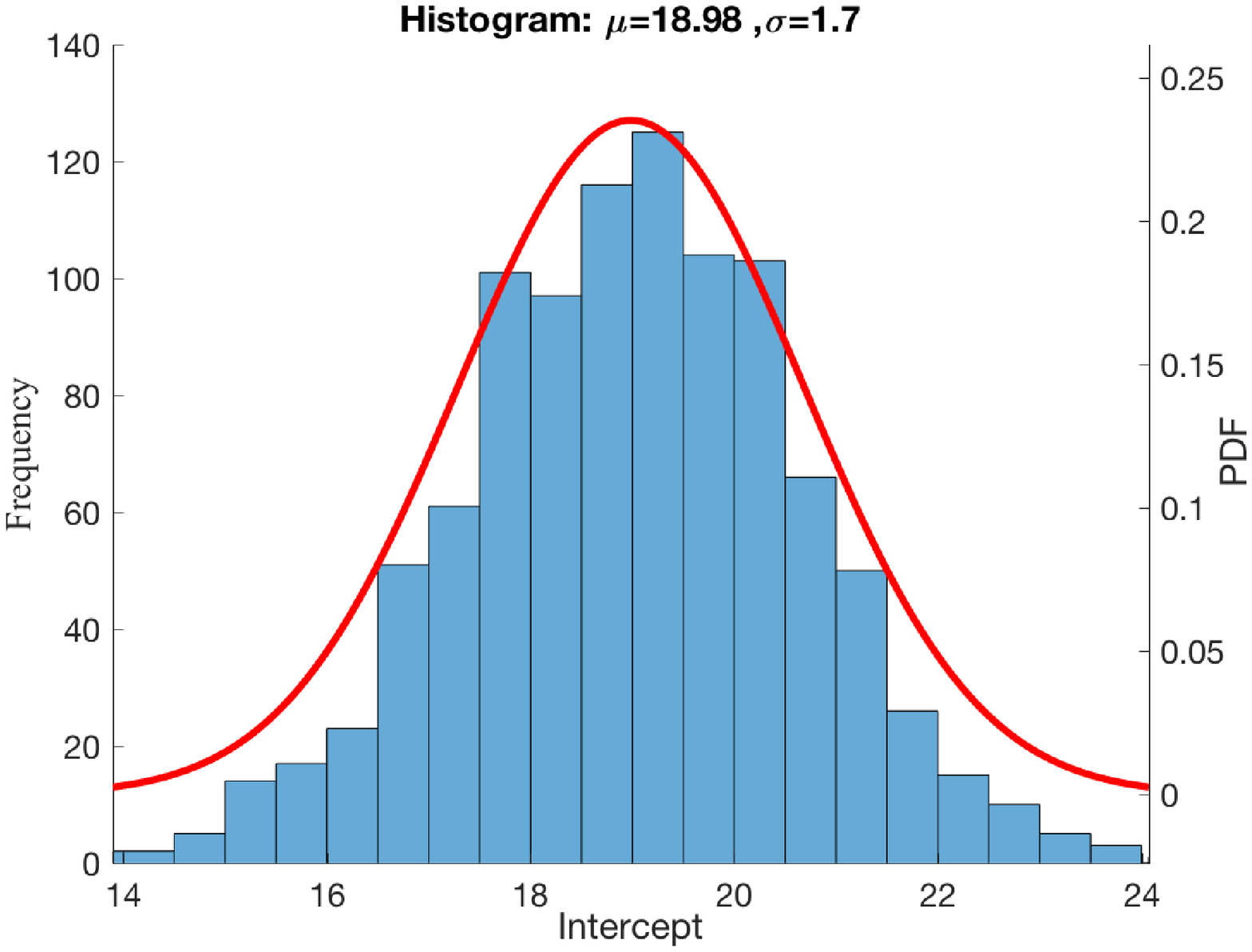}} 
	\subfigure[]{\includegraphics[width=\columnwidth]{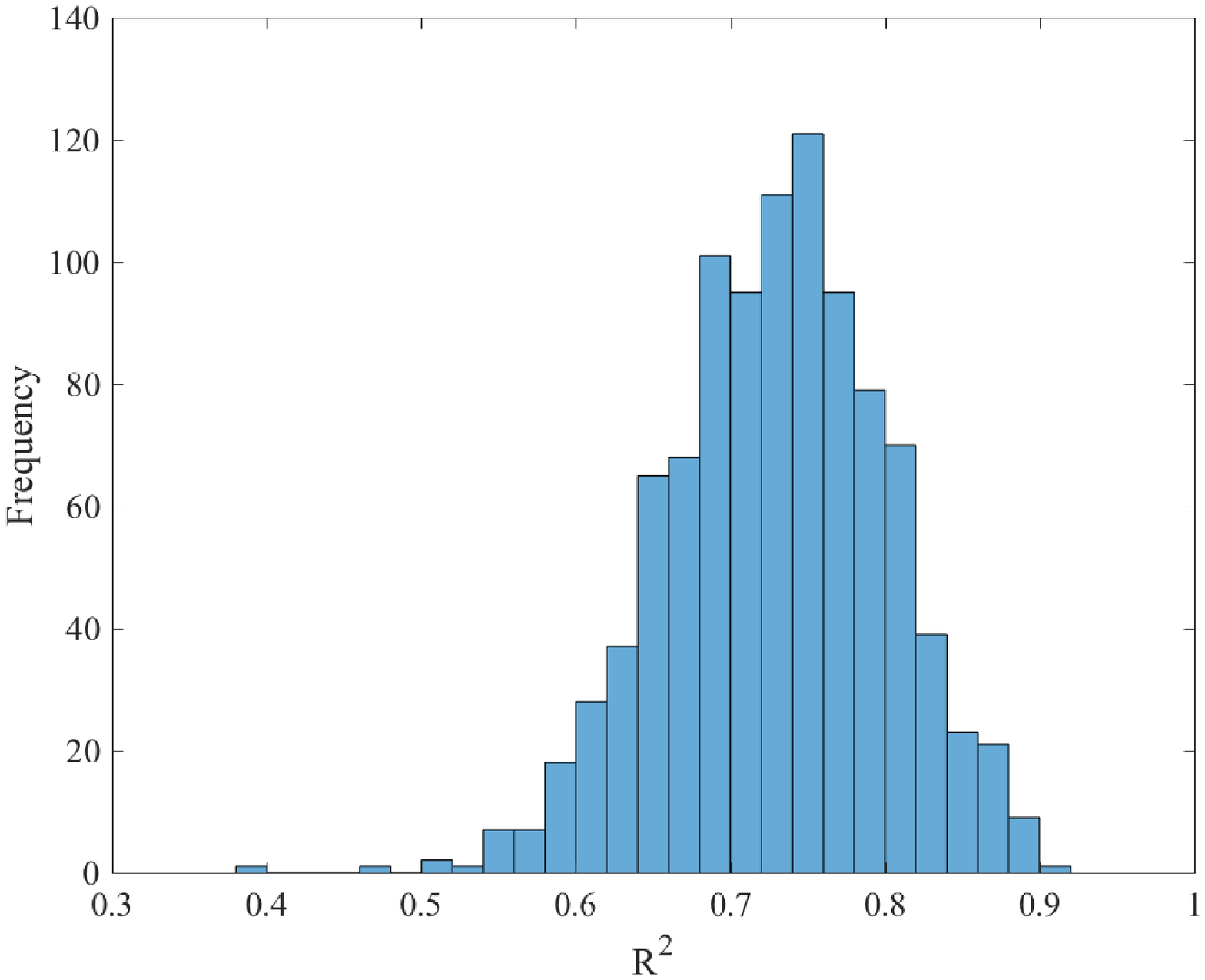}} \\

	\caption{(a) The extended 6.4 keV monochromatic luminosity versus log($L_{\rm OIV}$). Similarly, we run 1000 Monte Carlo simulations. The gray region represent the fitting results. (b) the corresponding linear slope distribution. Red line represents the fitted Gaussian model. (c) intercept distribution. (d) linear regression coefficient distribution.}
	\vspace{0.02in}
\end{figure*}

\subsection{Comparison with relevant Compton-Thick sources and Implications}

Our results on the extended Fe K$\alpha$ emission are consistent with recent findings in Compton thick AGNs, although the spatial extent and morphologies are diverse in the different cases. In NGC 1068 \citep{2015Bauer}, the Fe K$\alpha$ line has the extended components to the northeast and southwest in the direction of the ionization cone (\textgreater 140 pc). For Circinus \citep{2013Marinucci}, the excess in EW of Fe K$\alpha$ follows an axisymmetric geometry around the nucleus on hundred of pc scale. Moreover, part of the excess is co-aligned with the disk of the galaxy and almost perpendicular to the [OIII] ionization cone. In ESO 428-G014 \citep{Fabbiano2017Discovery}, the extended component with $\textgreater$ 2 kpc diameter follows the ionization cone. In NGC 4945 \citep{Marinucci2017}, the hard X-ray continuum and the Fe K$\alpha$ emission are confined to a flattened "torus" $\approx$ 200 $\times$ 100 pc region along the galactic disk perpendicular to the main axis of the ionization cone. As for NGC 5643 \citep{Fabbiano2018}, the Fe K$\alpha$ line feature is clumpy and extend for 65 pc perpendicular to the ionization cone. It is spatially consistent with the north-south elongation found in the high-resolution CO (2-1) imaging with ALMA, but slightly more extended than the rotating molecular disk of diameter $\approx$ 26 pc indicated by the kinematics of the CO line. Compared with the above Compton thick sources, NGC 4388 is the first Compton thin source(\nH $\simeq 3.6^{+0.27}_{-0.23}\times 10^{23}$ cm$^{-2}$ from this work), enabling a new perspective to investigate the properties of circumnuclear reprocessing matter in a different environment.

These results imply that Fe K$\alpha$ lines with large EW originated from circumnuclear reflector beyond the torus could be more common than previously thought. To obtain better understanding of these sources, we summarize their relevant properties in Table 4. Using the [OIV] 25.89 $\mu$m line luminosities as a robust indicator of AGN intrinsic power \citep{Mel_ndez_2008} and X-ray spectra observed by {\em XMM}-Newton/EPIC, \citet{2010Liu} examined the relationship between luminosity of the narrow Fe K$\alpha$ line  ($L_{\rm Fe}$) and $L_{\rm OIV}$. They compared between type I and type II AGNs of similar luminosities and found that statistically, the Fe K$\alpha$ luminosities in Compton-thin and Compton-thick type II AGNs are about 2.7 and 5.6 times lower than that in type I sources,  respectively.

Using six type II AGNs (Table 4), we can first verify the $log(L_{\rm [Fe K\alpha]tot})$-$logL_{\rm OIV}$ relation using linear regression. We also explore if there exists a similar linear correlation between extended Fe K$\alpha$ luminosity and the AGN intrinsic power (log$L_{\rm OIV}$). For Circinus galaxy, the whole extended Fe K$\alpha$ region is divided into six smaller regions \citep{2013Marinucci} and the total luminosity of extended Fe K$\alpha$ is their sum. For NGC 4945, \cite{Marinucci2017} select out four smaller circular regions only covered part of neutral Fe K$\alpha$ extended region. Thus, we have reprocessed the same data for NGC 4945 and obtained more precisely the luminosity of total extended Fe K$\alpha$ emission.

In order to take into account the uncertainties of data points, we adopt Monte Carlo simulations in the fitting process to evaluate the linear correlations. The errors for each data are assumed to follow Gaussian distribution. For each of the correlations ($log(L_{\rm [Fe K\alpha]tot})$-log($L_{\rm OIV})$ and $log(L_{\rm [Fe K\alpha]ext})$-log($L_{\rm OIV}$) ),  we run 1000 simulations and obtain the distributions of correlation coefficients (slope and intercept) and linear regression coefficient $R^{2}$ in Figures 9 and 10. The blue and green lines in Figure 9a represent the correlations obtained by \citet{2010Liu}  for Compton-thin and Compton-thick Seyferts. It is apparent that the correlation for Compton-Thin Seyfert 2 \citep{2010Liu} over-predicts log$L_{\rm Fe}$ for these sources. The red line in Figure 9a represents the best-fit correlation with the maximum $R^{2}$ in the simulations. We also obtained the range (3$\sigma$) of coefficients (slopes and intercepts) in Figures 9b and 9c, respectively. Adopting the 6.4 keV monochromatic luminosity as a proxy for the intensity of reflection spectrum, we also plotted the 6.4 keV monochromatic luminosity log($L_{Fe}/EW_{Fe}$) of extended regions versus log$L_{\rm OIV}$ in Figure 11 following the same method. 
For Circinus galaxy \citep{2013Marinucci}, the total 6.4 keV monochromatic luminosity (total reflective spectrum) is the sum of six smaller regions (i.e. $L_{\rm [Fe K\alpha]ext/EW_{[Fe]ext}}$$=$$\sum_{j=1}^N L_{\rm [Fe K\alpha]ext,j/EW_{[Fe]ext,j}}$, $N=6$). We have summarized the simulation results in Table 5. In all of the above mentioned fittings, the error propagation caused by division, addition and logarithm has been taken into account. The linear correlations shown in Figures 10 and 11 indicate that the Fe K$\alpha$ extended emission is connected with the central AGN other than alternative origin such as the unresolved X-ray binary emission. We may speculate that the scatter may be caused by the various covering factor, column density and clumpy distribution of reflector material around the nucleus.

Our findings may have implications on the detectability of extended Fe K$\alpha$ in Seyfert galaxies, no longer restricted to Compton-Thick AGNs only. Table 4 shows that the ratio of luminosity of extended Fe K$\alpha$ to total Fe K$\alpha$ can be up to 0.38 which indicates that the contribution from extended emission could not be neglected. There have been extensive studies of the molecular gas reservoir beyond the torus that can fuel the nuclear activity from surveys of nearby galaxies (e.g., the Molecular gas in NUclei of GAlaxies survey \citep{2003Ga,2005Garc}. If the geometry allows some of nuclear continuum to illuminate the dense gas available beyond the torus scale, similar reflection continuum and fluorescent Fe K$\alpha$ line could be produced.  Despite the relatively weak line luminosity compared to the central AGN, the EW can be large as seen here and in other cases. We speculate that if the central AGN ``changes look'' to a low state of inactive accretion, one would expect to find high fraction of extended Fe K$\alpha$ among local Seyfert galaxies where kpc scale structure can be resolved, which can be tested in future observations. 
	
	\section{Conclusions}
	
	We presented a detailed spectral and imaging analysis of the extended X-ray reflecting structure of Seyfert 2 galaxy NGC 4388, taking advantage of cumulative Chandra ACIS-S observations with 47.5 ks exposure.  The main results of this paper can be summarized as follows.
	
	\begin{enumerate} 
		
		\item Compared with previous results on NGC~4388 \citep{2003MN}, we focus on spatially resolved analysis of the kpc scale circumnuclear Fe K$\alpha$ emission with improved signal to noise ratio.  The spectral analysis reveals a spatial variation in the EW of Fe K$\alpha$ line for extended and nuclear regions by a factor of 3. The differences are likely to be associated with illuminated gas, in terms of angles between polar direction and the line of sight and the column density, but not attributed to a variation in metallicity. 
		
		\item The {\em HST} $V-H$ color map and IRAM CO (1-0) emission indicate that the extended Fe K$\alpha$ regions contain rich dust and gas.  The orientation of fluorescent Fe K$\alpha$ (north-east and west part) are along with the galactic disk.  Furthermore, the remaining extended regions appear to trace the edges of the radio jet, implying that the radio outflow in kpc scale may have compressed ambient interstellar gas, giving rise to enhanced line emission. 
		
		\item  We find that the luminosity of the extended Fe K$\alpha$ emission has a good linear correlation with the [OIV] luminosity which indicates that this emission is closely linked to the central AGN, instead of due to alternative origin such as unresolved X-ray binary emission. Using the 6.4 keV monochromatic luminosity ($L_{Fe}/EW_{Fe}$) for extended regions to represent the intensity of reflection continuum, we also find a good linear correlation with $L_{[OIV]}$.
		
	\end{enumerate} 
	
%	\section*{Acknowledgements}
	
	We thank the anonymous referee for her/his careful reading and suggestions that significantly improve our work. This work was supported by the National Key R \& D Program of China (2016YFA0400702) and the National Science Foundation of China (NSFC grants U1831205, 11473021, 11522323).
	X.S. acknowledges support from NSFC grant 11822301. C.P. work is supported by Fundação para a Ci\^{e}ncia e a Tecnologia (FCT) through the research grants PTDC/FIS-AST/29245/2017, UID/FIS/04434/2019, UIDB/04434/2020 and UIDP/04434/2020. H.Y. thanks Jie Zheng for assistance with data and software, Dr. Jiangtao Li, Xiaoyu Xu and Xiaodi Yu for discussion. We retrieved data from the NASA-IPAC Extragalactic Database (NED), the Hubble Space Telescope Archive and the Chandra Data Archive. For the data analysis, we used the CIAO, Sherpa and DS9, developed by the Chandra X-ray Center (CXC), and XSPEC developed by the HEASARC at NASA-GSFC. 
	
\facilities{Chandra (ACIS), XSPEC}

\software{CIAO (v4.9 \citep{2006SPIE.6270E..1VF}), Sherpa (\citep{2001SPIE.4477...76F}, \citep{2007ASPC..376..543D}),
XSPEC\citep{xs}, DS9 (\citep{2003}, Smithsonian Astrophysical
Observatory 2000), MARX \citep{2012SPIE.8443E..1AD} and ChaRT \citep{2003ASPC..295..477C} }
\\

	\bibliographystyle{apj}
	\bibliography{ms}

\end{document}